\documentclass{aa}

\usepackage{amsmath}

\usepackage{float}
\usepackage{txfonts}

\usepackage{hyperref}

\usepackage[usenames,dvipsnames,svgnames,table]{xcolor}
\definecolor{electricgreen}{rgb}{0.0, 1.0, 0.0}

\usepackage{appendix}
\usepackage[english]{babel}
\usepackage{comment}

\usepackage{graphicx}
\usepackage[caption = false]{subfig}

\usepackage{natbib}
\usepackage{multirow}
\usepackage{booktabs}

\usepackage{pdflscape}
\usepackage{lipsum, babel}

\begin{document}

\title{Polyaromatic disordered carbon grains as carriers of the UV bump: \\ FUV to mid-infrared spectroscopy of laboratory analogs}
\titlerunning{Polyaromatic disordered carbon grains as carriers of the UV bump}

\author{L. Gavilan\inst{1,2} 
          \and
          K. C. Le \inst{3}
         \and
          T. Pino\inst{3}
         \and
          I. Alata\inst{2,3}
          \and         
          A. Giuliani\inst{4,5}
        \and
          E. Dartois\inst{2}}

\institute{
Universit\'e Versailles St-Quentin; Sorbonne Universit\'es, UPMC Universit\'e Paris 06; CNRS/INSU, LATMOS-IPSL, 78280 Guyancourt, France
         \and
         Institut d\textsc{\char13}Astrophysique Spatiale (IAS), CNRS,  Univ. Paris Sud, Universit\'e Paris-Saclay, F-91405 Orsay, France
          \and
          Institut des Sciences Mol\'eculaires d'Orsay (ISMO), CNRS, Univ. Paris Sud, Universit\'e Paris-Saclay, F-91405 Orsay, France
          \and
          DISCO beam line, SOLEIL synchrotron, Saint Aubin, France 
          \and
          INRA, UAR1008 CEPIA, F-44316 Nantes, France\\
             \email{lisseth.gavilan@latmos.ipsl.fr}
           }

\abstract {A multiwavelength study of laboratory carbons with varying degrees of hydrogenation and sp$^2$ hybridization is required to characterize the structure of the carbonaceous carriers of interstellar and circumstellar extinction.} 
{We study the spectral properties of carbonaceous dust analogs from the far-ultraviolet to the mid-infrared and correlate features in both spectral ranges to the aromatic/aliphatic degree.} 
{Analogs to carbonaceous interstellar dust encountered in various phases of the interstellar medium have been prepared in the laboratory. These are amorphous hydrogenated carbons (a-C:H), analogs to the diffuse interstellar medium component, and soot particles, analogs to the polyaromatic component. Thin films (\textit{d} $<$ 100 nm) have been measured in transmission in the vacuum-ultraviolet (VUV; 120 - 210 nm) within the atmospheric pressure experiment (APEX) chamber of the DISCO beam line at the SOLEIL synchrotron radiation facility. Spectra of these films were further measured through the UV-Vis (210 nm - 1 $\mu$m) and in the mid-infrared (3 - 15 $\mu$m).}
{Tauc optical gaps, E$_{g}$, are derived from the visible spectra. The major spectral features are fitted through the VUV to the mid-infrared to obtain positions, full-widths at half maximum (FWHM), and integrated intensities. These are plotted against the position of the $\pi$-$\pi^*$ electronic transitions peak. Unidentified or overlapping  features in the UV are identified by correlations with complementary infrared data. A correlation between the optical gap and position of the $\pi$-$\pi^*$ electronic transitions peak is found. The latter is also correlated to the position of the sp$^3$ carbon defect band at $\sim$8 $\mu$m, the aromatic C=C stretching mode position at $\sim$6 $\mu$m, and the H/C ratio.}
 {Ultraviolet and infrared spectroscopy of structurally diverse carbon samples are used to constrain the nanostructural properties of carbon carriers of both circumstellar and interstellar extinction, such as the associated coherent lengths and the size of polyaromatic units. Our study suggests that carriers of the interstellar UV bump should exhibit infrared bands akin to the A/B classes of the aromatic infrared bands, while the circumstellar bump carriers should exhibit bands corresponding to the B/C classes.}
 

\maketitle

\section{Introduction}

The diversity of carbonaceous solids in interstellar and circumstellar environments in our Galaxy and in extra-galactic lines of sight has been the subject of  numerous theoretical, observational, and experimental studies  \citep{Draine1984, Greenberg1999, Pendleton2002, Dartois2005, Pino2008, Godard2010, Duley2012, Jones2012b, Jones2012a, Jones2012c}. Observations of interstellar extinction were first documented by \cite{Trumpler1930}. One prominent spectral feature in extinction curves in the Milky Way is a broad ultraviolet absorption bump consistently centered at 217.5 nm (4.6 $\mu$m$^{-1}$). The UV bump was first observed in the 1960s by \cite{Stecher1965}, but its carrier is still not fully understood, although it has been generally attributed to electronic transitions in carbon. \cite{Fitzpatrick1986} first parametrized the UV bump using data from the International Ultraviolet Explorer (IUE) toward 45 reddened OB Milky Way stars. They found that the position of the bump was mildly variable while its full width at half maximum strongly variable; the lack of correlation between these parameters suggested that graphite was an improbable carrier. In \cite{Fitzpatrick1988} they included the far ultraviolet (FUV)  extinction, finding that a Lorentzian (for the UV bump), a FUV rise, and an underlying linear continuum were necessary components. This study was expanded to 78 stars, published as an atlas \citep{Fitzpatrick1990}, and further extended to include 329 stars \citep{Fitzpatrick2005, Fitzpatrick2007}. In the latter,  stellar atmosphere models were used to eliminate spectral mismatch errors in the curves, and showed that extinction parameters are highly sensitive to local conditions. The exact origin and variability of interstellar extinction and the nature of the dust carriers across different wavelength ranges is still a central question in cosmic dust research. 
 
Astronomical observations in the infrared also play an important role in dust characterization.  
Unidentified infrared emission bands (UIBs), also known as aromatic infrared bands (AIBs), are features between 3 and 20 $\mu$m (3.3, 6.2, 7.7, 8.6, and 11.2 $\mu$m) widely observed in a range of environments in our Galaxy \citep[e.g.,][]{Rosenberg2014, Peeters2014}. These bands are attributed to the infrared fluorescence of polyaromatic carriers, associated to the polycyclic aromatic hydrocarbon (PAH) hypothesis \citep{Leger1984, Puget1985, Allamandola1985, Puget1989, Allamandola1989}. However, alternative carriers have also been proposed, such as amorphous organic solids with mixed aromatic-aliphatic structures due to mixed sp$^2$ and sp$^3$ hybridizations. Observations of a single carbon star with a circumstellar disk by \cite{Sloan2007} concluded that its spectra was consistent with hydrocarbon mixtures containing both aromatic and aliphatic bonds. \cite{Acke2010} suggested that a strong UV flux reduces the aliphatic component and emphasizes the spectral signature of the aromatics in the emission spectra towards Herbig Ae stars. \cite{Yang2013} studied the ratio of the CH stretching modes at 3.3 and 3.4 $\mu$m in emission towards UIBs, and argued that their observed emitters are predominantly aromatic nanoparticles.  

The spectroscopy of laboratory analogs to interstellar carbon dust has helped to unveil the origin of the UV bump. \cite{Draine1984} proposed early on that crystalline graphite could fit the UV bump, and later on  \cite{Jones1990} argued that hydrogenated amorphous carbons (a-C:H) containing small polyaromatic units may be suitable carriers. Observations of the aliphatic bands at 6.85 and 7.25  $\mu$m in absorption later revealed the presence of interstellar hydrogenated amorphous carbons \citep{Pendleton2002, Dartois2004}. Laboratory hydrogenated carbons have since then been considered realistic carriers of the bump \citep{Colangeli1997, Schnaiter1998, Furton1999, Dartois2004}. Laboratory a-C:Hs at  different degrees of hydrogenation have been compared to observed astronomical spectra \citep{Dartois2007, Duley2012}. Soot nanoparticles prepared via laser pyrolisis \citep{Jager2006, Jager2008} are also candidates but only following the activation of $\pi$-$\pi^*$ transitions by UV irradiation \citep{Gadallah2011}. Indeed, energetic thermal and non-thermal processes can alter carbonaceous dust in the diffuse interstellar medium (ISM). In photo-dissociation regions, photodesorption processes compete with hydrogenation to effectively alter their structure and optical properties. This has been experimentally explored via UV irradiation of thin a-C:H films \citep{Scott1996, Mennella1996, Mennella2002, Alata2014, Alata2015}, as well as via ion irradiation \citep{Mennella2003, Godard2011}. Recent optical models of interstellar dust \citep{Jones2012b, Jones2012a, Jones2012c,Jones2013} have shown that amorphous carbons extending over a wide range of optical gap energies can be used to fit the observed extinction from the FUV to the infrared. 
 \cite{Russo2014} performed a UV study extending from 190 to 1100 nm on molecular-weight-sorted species prepared by combustion. \cite{Bescond2016} recently studied the optical properties of soots in the visible near-UV from 200-1100 nm and performed an inversion model to extract the size distribution and fractal dimension of soots. Few laboratory spectroscopic studies on carbonaceous dust extending over a wide spectral range exist, due to experimental difficulties in accessing the vacuum-ultraviolet (VUV). In this range, different techniques have been used including electron energy loss spectroscopy (EELS)  \citep{Fink1984, Zubko1996} and synchrotron VUV sources \citep{Colangeli1993, Gavilan2016a}.

For this study, we produced in the laboratory carbonaceous solids of mixed aromatic and aliphatic structures at different degrees of hydrogenation. By measuring them over a large spectral range (from the mid-infrared to the VUV), we expect to unveil relations between vibrational modes and electronic transitions, both observable signatures of interstellar dust, extending our previous findings in \cite{Gavilan2016a}. We aim to constrain the carbonaceous carriers of the interstellar UV bump and the circumstellar (CSM) UV bump and shed light on the spectroscopic decomposition of UV extinction based on our laboratory data. For the ISM, our work will be compared to the parameters of the interstellar 217.5 nm UV bump \citep{Fitzpatrick2005, Fitzpatrick2007}. For the CSM, our work will be compared against the very few UV observations of circumstellar envelopes, in this case of H-poor post-asymptotic giant branch (AGB) and (hot) R Cor Borealis stars \citep{Drilling1997, Blanco1995, Waelkens1995}. Although this CSM is not representative of H-rich carbon stars, which provide most of the galactic carbonaceous dust, we use these UV observations as a first constraint to the circumstellar UV bump.

This article is organized as follows. In Section \ref{sec:2} we describe the experimental setups and sample preparation. In Section \ref{sec:3} we present the analysis of UV and IR bands. In Section \ref{sec:dis} we discuss our analysis and astronomical implications, and in Section \ref{sec:5} we present our conclusions. 

\section{Experiments}
\label{sec:2}

\begin{figure*}[htbp]
\centering
\caption{Tauc optical gap extrapolation using the visible spectra for soot samples prepared with different C/O flame ratios and collected at varying heights above the burner (HAB in mm). }
\vspace{1em}
\subfloat[ Combustion flame with C/O = 0.82 ]{\includegraphics[width = 86mm]{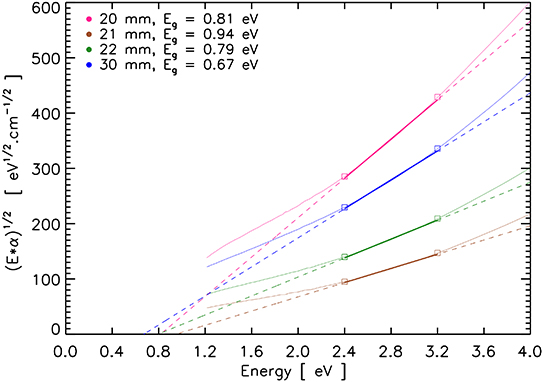}} \hspace{1cm}
\subfloat[ Combustion flame with C/O = 1 ]{\includegraphics[width = 86mm]{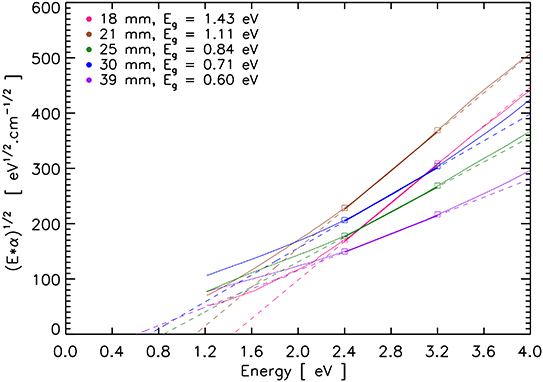}}\\ 
\subfloat[ Combustion flame with C/O = 1.05 ]{\includegraphics[width = 86mm]{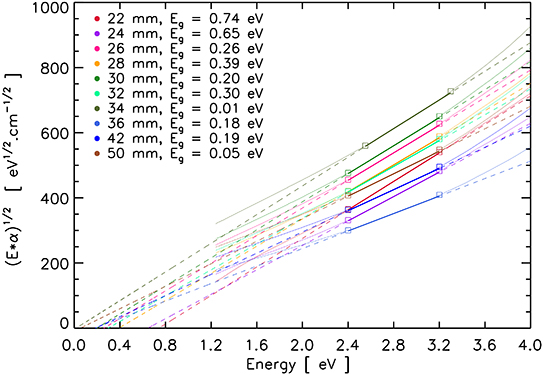}} \hspace{1cm}
\subfloat[ Combustion flame with C/O = 1.3 ]{\includegraphics[width = 86mm]{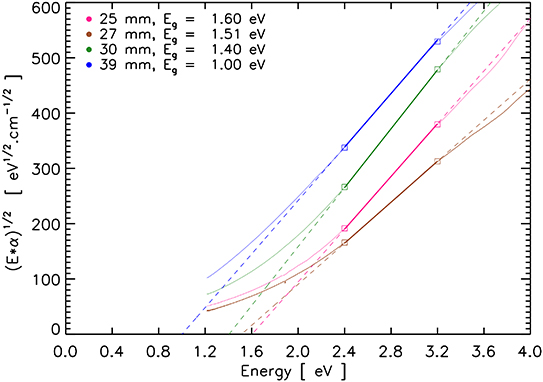}}
\label{fig:egap}
\end{figure*}

The experimental setups used to prepare the carbonaceous samples have already been described in \cite{Gavilan2016a}. Here we provide further details relevant to the samples used in this study. Thin carbon films ($d$ $<$ 100 nm) deposited on MgF$_2$ windows were used for measurements between 120 - 1000 nm, and thicker carbon films ($d$ $>$ 300 nm) deposited on IR transparent KBr, CsI, and KCl windows for measurements in the mid-infrared. 

\textit{SICAL-ICP}, a low pressure ($\sim$10$^{-2}$ mbar) radiofrequency (R.F.) plasma reactor was used to prepare the hydrogenated amorphous carbon samples. A 13.56 MHz R.F. power is inductively coupled to the plasma by impedance matching via a copper coil surrounding a glass cylinder \cite[e.g.,][]{Dworschak1990}. While the a-C:H sample presented in \cite{Gavilan2016a} was prepared using methane (CH$_4$) as the precursor gas, and an R.F. input power of 100 W, the a-C:H presented in this paper was prepared using a plasma fed by  toluene (C$_7$H$_8$), and an R.F. input power of 150 W. Our goal in this study was to produce a more aromatic a-C:H. The higher R.F. power (150 versus 100 W) has an important influence on the structure of the produced solid hydrocarbon. Higher ion energies lead to preferential sputtering of hydrogen due to the lower binding energies of CH with respect to CC bonds. 

We used the \textit{Nanograins} combustion flame setup to produce soot nanoparticle films \citep{Carpentier2012, Pino2015}. In premixed flames, sooting is the result of competition between the rate of pyrolysis, growth of soot precursors, and oxidation rate \citep{Mansurov2005}. The formation of soot nanoparticles can be described by the following equation, 
\begin{gather}
\label{eq1}
{\rm C_m H_n + \alpha O_2 \rightarrow 2 \alpha CO + \frac{n}{2}H_2 + (m-2\alpha)C_s },
\end{gather} 
where $\mathrm{m}$  and $\mathrm{n}$ are positive integers related to the hydrocarbon gas  precursor type, $\mathrm{s}$ is a positive integer related to the number of carbon atoms in a solid soot nanoparticle, and $\mathrm{\alpha}$ is a positive rational for the oxidation coefficient. Theoretically, for soots to form in an ethylene flame, Eq. \ref{eq1} imposes C/O $>$ 1. In practice, the sooting regime appears as soon as C/O $\sim$0.7 at 40 mbar. Soots produced in an ethylene (C$_2$H$_4$) flame, first introduced in \cite{Gavilan2016a}, were prepared under the following conditions: C/O ratio = 1.05, height above the burner (HAB) = 30 mm. In this paper we present soots prepared under varying flame conditions and HAB ranging from 18 to 50 mm.  We used the following C/O ratios and pressures: 0.82 (60 mbar), 1 (28 mbar), 1.05 (33 - 40 mbar), 1.3 (40 mbar). The deposition time of the films varied from approximately one to four minutes for the closest HAB of 25 mm to 1 to 20 seconds beyond 30 mm. 

\section{UV and IR spectroscopy}
\label{sec:3}

\begin{figure}[htbp!]
\begin{center}
\includegraphics[width=86mm]{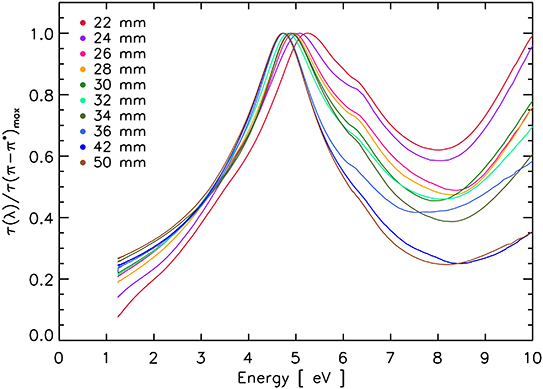}
\caption{VUV-Visible spectra of a series of soot nanoparticle films (C/O = 1.05, HAB = 22 - 50 mm). These have been normalized by the maximum optical depth of their respective $\pi$-$\pi^*$ electronic band.}
\label{fig:vuvas}
\end{center}
\end{figure}

VUV spectra of all samples were recorded in the 120 - 250 nm range using the APEX branch of the DISCO (dichroism, imaging and spectroscopy for biology) beam line \citep{Giuliani2009} at the SOLEIL synchrotron. These spectra were completed in the UV-Vis (210 - 1000 nm) measured with a spectrophotometer (Specord, \textit{Analytikjena}). The overlapping energy range was used to combine the VUV and UV-Vis datasets.
Infrared spectra of the same films were measured using an evacuated fourier transform infrared (FTIR) spectrometer (Bruker Vertex 80V) equipped with a KBr beam-splitter and a HgCdTe (mercury cadmium telluride, MCT) detector working in the 4000 to 600 cm$^{-1}$ (2.5 to 15 $\mu$m) at 2 cm$^{-1}$ resolution. 
The infrared spectra were corrected with a polynomial baseline to subtract the continuum absorption increasing towards higher wavenumbers due to electronic transitions that extend into the infrared. The spectroscopic analysis involved deconvolution of the main electronic bands for the VUV-UV spectra and for the vibrational modes between 3300 and 600 cm$^{-1}$ for the infrared spectra using the IDL MPFIT routine based on the Levenberg-Marquardt non-linear least-squares minimization method \citep{Markwardt2009}. The VUV spectra of a sample soot series are shown in Fig. \ref{fig:vuvas}. All VUV-MIR spectra and the respective deconvolved features are included in appendix \ref{app1} of the electronic version.

\subsection{Relationships among the UV fit parameters}
\label{sec:uvfit}
UV-VUV spectra with enough S/N were selected and deconvolved using a Lorentzian and three Gaussian functions. We called these the L1, G1, G2, and G3 bands. L1 is the main UV absorption band, falling between 190 to 260 nm, and it is in most cases well-fitted by a single Lorentzian. G1 and G2 are neighboring peaks, appearing around 300 nm and 190 nm respectively. G3 is the large UV absorption at around 110 nm. No additional linear background was required for these fits. The resulting fit UV-VUV parameters are summarized in Table \ref{Table1} of the electronic version.

\subsubsection{Determination of the optical gap}

We used the visible spectral range (350 - 1000 nm) to determine the optical band gap of these materials, also known as the Tauc gap. This is often used to characterize the optical properties of amorphous semiconductors \citep{Tauc1966}. The Tauc gap is determined only by the $\pi$-states (mostly or totally delocalized) and is intrinsically related to the energy gap between the valence and conduction bands \citep{Silva1998}. The Tauc gap has also been used to generalize the optical properties of interstellar amorphous carbons \citep{Duley1992, Jones2012b}. For our samples, the Tauc gap values were obtained by extrapolating the linear region in an ($\alpha$$\times$E)$^{1/2}$ versus E plot, where $\alpha$ is the absorption coefficient obtained from the relation $\alpha$ = $\tau$/h, where $\tau$ is the optical depth, \textit{h} the film thickness, and E is in eV.
For disordered carbon materials like those presented in this paper, the empirical linear dependance between $\alpha$ and E is expressed by,
\begin{gather}
(\alpha \times E)^{1/2} = B^{1/2}(E-E_g),
\end{gather}
where $B$ is a constant attributed to their thermal processing \citep{Jones2012b}. To determine the Tauc gap from the visible-ultraviolet spectra, we chose the same region in the visible tail of the Tauc spectra where 2.4 $<$ E (eV) $<$ 3.2 to extract the linear fit extrapolated to zero absorption, and from 2.4 - 2.7 eV and 2.7 - 3.2 eV to determine the average systematic error bar typically used in soot extinction analysis \citep{Adkins2015}. For the soot produced with the C/O = 1.05 flame, HAB = 34 mm,  the fitting linear energy range was shifted to 2.5 - 3.3 eV, in order to avoid a negative gap value. While the thickness of the samples is assumed to be $\sim$30-60 nm as in \cite{Gavilan2016a}, this is taken as constant for all samples as it does not significantly affect the gap value determination, whose systematic uncertainty is dominated by the choice of linear energy range. Figure \ref{fig:egap} shows the calculated gap values for each film. For the prepared samples, the value of the Tauc optical gap (E$_{g}$) increases from the least hydrogenated soots to the most hydrogenated a-C:H, that is, 0 $<$ E$_g$ $<$ 2 eV. The H/C ratio of each sample is obtained from infrared spectra and discussed in Section \ref{sec:ir}.   

For the a-C:H prepared with a toluene precursor, the optical gap was found to be 1.95 eV, lower than the 2.5 eV estimated for the methane-precursor a-C:H of \citep{Gavilan2016a}. For the soot samples produced with the C/O = 0.82 flame, E$_g$ decreases from 0.78 to 0.62 eV as the HAB increases. This is the same trend as for the C/O = 1 flame, where E$_g$ decreases from 1.46 to 0.68, for the C/O = 1.05 flame with a decrease from 1.47 to 0.01, and for the C/O = 1.3 flame with a decrease from 1.6 to 1 eV.  
\cite{Jones2012b} sequences the family of amorphous carbons based on the optical gap energy, building on previous work by \cite{Robertson1986}. Amorphous carbons that are hydrogen rich, called a-C:H, have E$_g$ $\simeq$ 1.2 - 2.5 eV, while the hydrogen poor carbons, called a-C, have E$_g$ $\simeq$ 0.4 - 0.7 eV. For our soot samples E$_g$ $<$ 1.6 eV, and E$_g$ = 1.95 for our a-C:H. Thus, Robertson's classification can be further improved by our disordered carbon data.

\subsubsection{E$_{g}$ versus $\omega_c$(L1)}

\begin{figure}[htbp]
\begin{center}
\includegraphics[width=90mm]{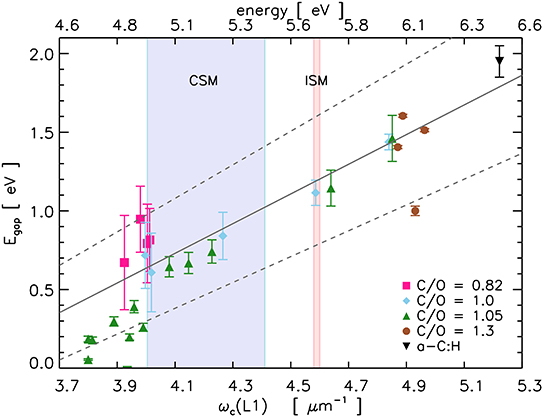}
\caption{Tauc optical gap, E$_{g}$, for our laboratory samples versus the center of the $\pi$-$\pi^*$ electronic transition, $\omega_c$(L1). The latter is used as our main spectroscopic parameter in the UV.}
\label{fig:Eg}
\end{center}
\end{figure}

In Fig. \ref{fig:Eg} we display the position of the main UV absorption (fitted by a Lorentzian, called L1) against the optical gap, yielding the relation,
\begin{gather}
E_{g} [ eV ] = 0.76 \pm 0.3 \times \omega_c(L1) [ \mu m^{-1} ] - 3.1 \pm 0.5.
\end{gather}
These two parameters have a linear Pearson correlation coefficient \textit{r} = 0.91.  The wide variation in the position of the L1 peak from 3.8 to 5.2 $\mu$m$^{-1}$ (192 to 262 nm) is  consistent with previous experimental works where the UV peak is centered between 180 and 280 nm \citep{LlamasJansa2003}. \cite{Russo2014} found that the spectral features in the UV-visible range are mainly affected by the higher molecular weight species of combustion flame products.  
Because L1 is strongly correlated to the optical gap E$_g$, variations in the position of L1 can be attributed to changes in the nanostructuration due to varying H/C, sp$^3$/sp$^2$ , and the size of the polyaromatic units, which will be discussed in the following sections. The Tauc optical gap model, which has been used to give a complete description of the optical properties of carbonaceous materials \citep{Robertson1987, Robertson1991, Jones2012b}, appears to be well-suited for our soot samples. 

From now on, our fitting parameters will be compared to the parameters of the interstellar 217.5 nm UV bump \citep{Fitzpatrick2005, Fitzpatrick2007}, that is, $\omega_c$ = 4.593 $\mu$m$^{-1}$ with $\sigma$ = 0.0191 $\mu$m$^{-1}$  and FWHM = 0.77 - 1.29 $\mu$m$^{-1}$ with $\sigma$ = 0.050 $\mu$m$^{-1}$. We also compare our parameters to the average position of the circumstellar $\sim$240 nm UV bump, as determined by observations of the circumstellar envelopes of C-rich stars like V348 Sagittarii \citep{Drilling1997}. For the average circumstellar UV  bump, $\omega_c$ = 4.12 $\mu$m$^{-1}$, limited from 4 $\mu$m$^{-1}$ \citep{Blanco1995} up to 4.4 $\mu$m$^{-1}$ \citep{Waelkens1995}, and a FWHM = 1.34 $\mu$m$^{-1}$ with $\sigma$ = 0.06 $\mu$m$^{-1}$ \citep{Drilling1997}.
The spectral parameters of our samples will be preferably plotted against the position of L1, $\omega_c$(L1), strongly correlated to E$_g$, as L1 (rather than E$_g$) is the direct spectroscopic parameter. 

\subsubsection{$\gamma$(L1) versus $\omega_c$(L1)}

\begin{figure}[htb!]
\begin{center}
\includegraphics[width=90mm]{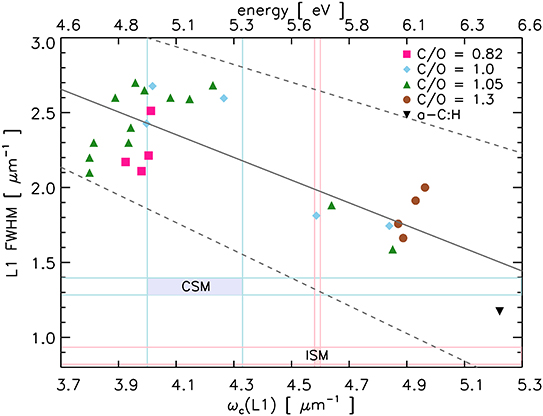}
\caption{FWHM of the $\pi$-$\pi^*$ electronic transition, $\gamma$(L1), versus its center position, $\omega_c$(L1), showing a clear correlation, absent in interstellar extinction curves \citep{Fitzpatrick2005}.}
\label{fig:cor1}
\end{center}
\end{figure}

As seen in Fig. \ref{fig:cor1}, an anti-correlation is found between the FWHM $\gamma$(L1) and $\omega_c$(L1) (\textit{r} = -0.79), while $\gamma$(L1) versus E$_{g}$ is also anti-correlated (\textit{r} = -0.72). This result  contrasts with the non-correlation between both parameters in extinction curves found by \cite{Fitzpatrick2005}. The large scatter in FWHM has been suggested to be due to grain clustering  broadening \citep{Schnaiter1998}. From all our samples, only one has an L1 peak within the observed ISM FWHM parameter range, the a-C:H prepared with toluene, with $\gamma$(L1) = 1.2 $\mu$m$^{-1}$, but it is centered at 5.2 $\mu$m$^{-1}$ and not at the expected 4.6 $\mu$m$^{-1}$.  However, if we consider a different parametrization of the interstellar UV bump (Sect. \ref{sec:dis}), two soot samples are more representative, C/O = 1 (HAB = 21 mm ) and C/O = 1.05 (HAB = 18 mm). 

\subsubsection{$\omega_c$(G1, G2, G3) versus $\omega_c$(L1)} 

\begin{figure}[htbp]
\begin{center}
\includegraphics[width=90mm]{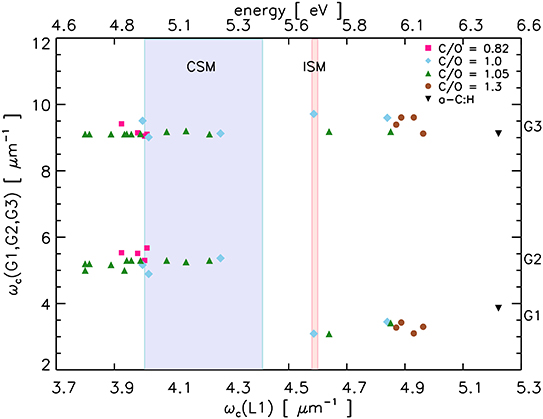}
\caption{Positions of the UV and VUV major electronic transitions, $\omega_c$(G1), $\omega_c$(G2), $\omega_c$(G3), versus the center of the $\pi$-$\pi^*$ electronic transition $\omega_c$(L1). For all samples presented here the positions of these peaks fall at precise energies, with little dispersion, but their relative intensities tend to vary strongly.}
\label{fig:cor3}
\end{center}
\end{figure}

In Fig. \ref{fig:cor3} we plot the positions of the main electronic transitions measured in the UV and VUV. G1 and G2 are absorption peaks in the neighborhood of L1. We can see that each electronic transition falls within a particular energy range for all samples: G1 peaks are constrained between 4.8 and 5.7 $\mu$m$^{-1}$ while G2 peaks fall between 3 and 4 $\mu$m$^{-1}$. The G1 and G2 transitions rarely appear simultaneously. The position of the G3 peak falls consistently at an average of 9 $\mu$m$^{-1}$. The ratio of the integrated G3 band to the L1 feature averages 2.0 $\pm$ 0.8. 

\subsection{Relationships among the UV and mid-IR fit parameters}

To complement the UV and VUV fit parameters, a detailed band analysis of the C-H stretching mode region (3100-2700 cm$^{-1}$) and bending mode region (1800-500 cm$^{-1}$)  was performed. The out of plane (OOP) bending modes in the 900 to 700 cm$^{-1}$ (11 to 15 $\mu$m) are generally used for the classification of the aromatic ring edge structures \citep{Russo2014}, but these are only present in the soots prepared in flames with C/O = 0.82 and 1.05. The mid-infrared attributions are found in previous soot studies \citep{Carpentier2012, Dartois2007}. Generally, the bending mode region is $\sim$1.5-4 times more absorbing than the stretching mode region for the soot samples. In a few cases, the integrated intensities of these modes were normalized to the most intense mode of the respective region for intercomparison between samples, to account for thickness variations between samples. The resulting fitted mid-infrared parameters are summarized in Table \ref{Table2} of the electronic version.

\subsubsection{$\omega_c$(sp$^3$ defect) versus $\omega_c$(arom. C=C stretch.)}

\begin{figure}[htbp]
\begin{center}
\includegraphics[width=90mm]{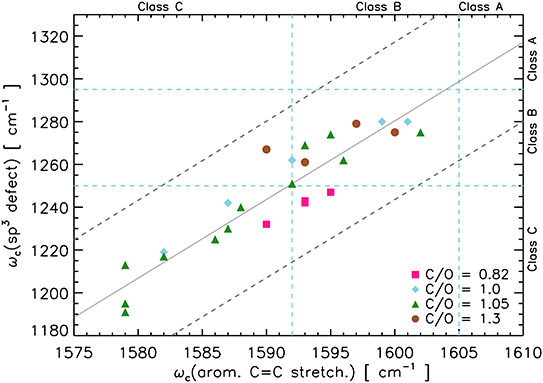}
\caption{Correlation between the carbon defect band, $\omega_c$(sp$^3$ defect), versus the aromatic C=C stretching band, $\omega_c$(C=C), verifying previous findings in other soot samples \citep{Carpentier2012}.}
\label{fig:wsp-wcc}
\end{center}
\end{figure}

In Fig. \ref{fig:wsp-wcc}, we show the strong correlation (r$^2$ = 0.89) between the $\omega_c$(aromatic C=C stretch) and $\omega_c$(sp$^3$ defect) band positions, confirming  previous findings \citep{Carpentier2012}. This relation traces differences in shapes and structures of the polyaromatic units in the soot. For polyaromatic carbons, three astrophysical AIB classes are proposed based on their spectra in the 6 to 9 $\mu$m$^{-1}$ region \citep{Peeters2002}. According to this classification, our laboratory soots correspond to classes B and C of the AIBs. 

\subsubsection{$\omega_c$(sp$^3$ defect) \&  $\omega_c$(arom. C=C stretch.) versus $\omega_c$(L1)}

\begin{figure}
\centering
\vspace{1em}
\subfloat[ ]{\includegraphics[width = 90mm]{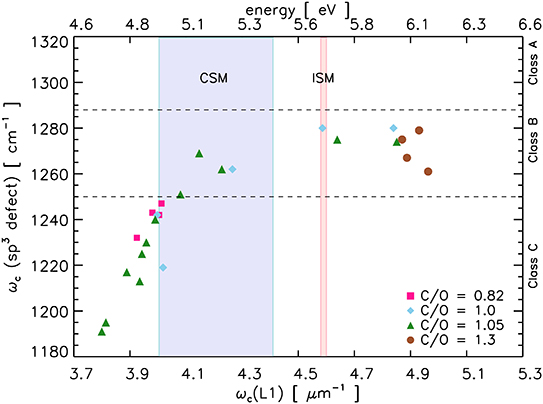}}\\ \vspace{-1em}
\subfloat[  ]{\includegraphics[width = 90mm]{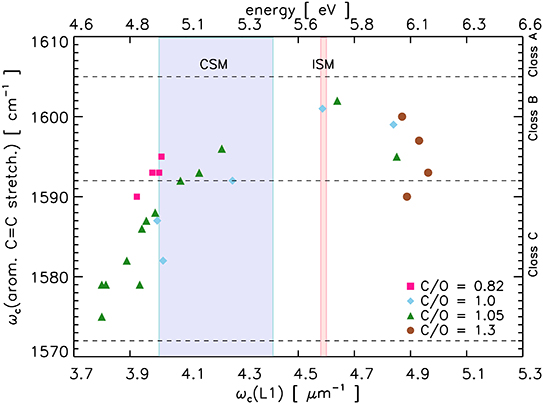}}  
\caption{Aromatic infrared bands (AIBs) in our laboratory soot samples. (a) $\omega_c$(sp$^3$ defect) versus $\omega_c$(L1). (b) $\omega_c$(arom. C=C stretch.) versus $\omega_c$(L1). The laboratory found band positions for the sp$^3$ defect and aromatic C=C band are compared to the classification of AIBs by \citep{Peeters2002}. Our laboratory samples are classified as Class B and Class C AIBs. The inferred CSM UV bump carriers fall within class B/C, and the ISM UV bump carriers approaches the A/B class.}
\label{fig:aib1}
\end{figure}

The sp$^3$ defect band is assigned to edge defects that can occur due to aliphatic cross-linking and as non-hexagonal rings carbon defects \citep{Carpentier2012} and contribute to electronic intra-band states of the material. For our samples, this band appears between 1190 - 1280 cm$^{-1}$. 

In Fig. \ref{fig:aib1}(a) we plot the position of the sp$^3$ defect band versus the center of the L1 peak, which correlates well for all of our samples. From this relation, the L1 position of the CSM bump would correspond to an sp$^3$ defect located $\sim$1260 cm$^{-1}$ (i.e., class B/C carrier), while for the ISM bump we expect it to be found $\sim$1280 cm$^{-1}$ (i.e., class B carrier). 

In Fig. \ref{fig:aib1}(b) we plot the position of the aromatic C=C band versus the center of the L1 peak, also well correlated, as expected from the relation found in Fig. \ref{fig:wsp-wcc}. For the a-C:H, the C=C band is found at 1625 cm$^{-1}$ and for the soot with the highest E$_g$ it is found at 1590 cm$^{-1}$. As the value of E$_g$ decreases (and the H/C ratio of the sample decreases), the position shifts down to 1575 cm$^{-1}$.  
The correlation is linear up to L1 positions $\sim$4.7 $\mu$m$^{-1}$. At higher energies this correlation shows a larger scatter. For the L1 position of the CSM bump, we would expect the C=C vibrational mode to be located at $\sim$1592 cm$^{-1}$ while for the ISM bump, we expect the defect to be found near 1600 cm$^{-1}$. 
It is then not surprising that the sp$^3$ defect and C=C band positions correlate with the position of L1 as they are all sensitive to the electrons' delocalizations in sp$^2$ conjugated systems. 

\subsection{Origin of UV features G1 and G2}

\subsubsection{ A(G1)/A(C=O) versus $\omega_c$(L1)}

G1 is a minor peak appearing for some soots (C/O = 1.3, for all distances from the burner, C/O = 1.05 for 14 and 18 mm) and for the a-C:H. To identify the origin of the G1 peak present in some UV spectra (for flames with C/O ratios = 1, 1.05 and 1.3), we plotted the Lorentzian position versus the ratio of the areas of A(G1) to A(C=O), showing that this feature appears for a certain type of carbon, whose main $\pi$-$\pi^*$ band appears at $\omega_c$ (L1) $>$ 4.6 $\mu$m$^{-1}$. 

We choose the thickest films, where the carbonyl mode is clearly observed, to identify the origin of G1. These samples correspond to flames with C/O = 1, 1.05. The flame with C/O = 1.3  produced very thin films where oxidation is surface dominated. By plotting the ratio of the G1 feature to the L1 feature in the UV to the ratio of the integrated C=O to the C=C mode, a weak correlation arises. By following the evolution of the mid-infrared and UV spectra for the flame with C/O = 1 in  the series of Fig. \ref{fig:a2}, as soon as the intensity of C=O $\ll$ C=C, the UV G1 peak disappears. The G1 peak can be assigned to the oxidation of the younger soots produced near the burner, as traced by the C=O infrared intensities. 
 
\subsubsection{A(G2)/A(CH$_{arom.}$) versus $\omega_c$(L1)}

G2 is a minor peak appearing for some soots (C/O = 1, beyond 25-39 mm, C/O=1.05, beyond 22 mm) at $\sim$5.3 $\mu$m$^{-1}$ (180 nm). This peak appears as the soots are collected farther from the burner, that is, when they are more mature soots with polyaromatic structures or fullerene-like soots \citep{Carpentier2012}. \cite{Gadallah2011} and \cite{LlamasJansa2003} attributed this peak to carbonyl vibrations. We propose a different attribution. We divided the integrated area of G2 by the integrated area of the aromatic =CH stretching mode, and confirm that G2 appears only when the intensity of the sp$^3$ defect $\gg$ C=C band and for $\omega_c$ (L1) $<$ 4.3 $\mu$m$^{-1}$, corresponding to the lower Tauc gap materials, that is, low H/C soots where defects abound.  

\subsection{H/C ratios}
\label{sec:ir}

\begin{figure}[htbp]
\begin{center}
\includegraphics[width=90mm]{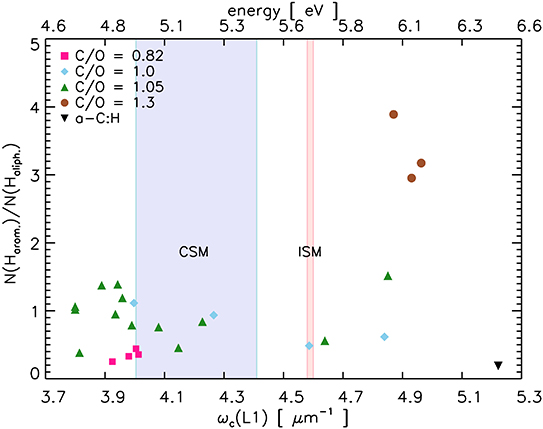}
\caption{Aromatic to aliphatic hydrogen ratio versus $\omega_c$(L1), determined from the integrated infrared =CH and -CH$_{2,3}$ stretching modes, corrected by the ratio of the corresponding oscillator strengths.}
\label{fig:hc1}
\end{center}
\end{figure}

The mid-infrared spectra of different carbonaceous samples reveal the carbon hybridization states and allow us to discriminate between aliphatic and aromatic hydrogen. The C-H stretching peaks in the 3100-2900 cm$^{-1}$ allow us to determine the aromatic to aliphatic hydrogen ratio. We take the ratio of the integrated CH$_{aromatic}$ ($\sim$3040 cm$^{-1}$) to the CH$_{aliphatic}$ modes (CH$_2$ asymmetric stretching mode, $\sim$2920 cm$^{-1}$ including the Fermi resonance and the CH$_3$ asymmetric mode at $\sim$2960 cm$^{-1}$) normalized by their oscillator strengths, A(CH$_{arom}$) = 1.9 $\times$ 10$^{-18}$ cm/group, A(CH$_{2,as.}$) = 8.4 $\times$ 10$^{-18}$ cm/group, and A(CH$_{3,as}$) = 1.25 $\times$ 10$^{-17}$ cm/group \citep{Dartois2007}. With the exception of a single sample (C/O = 1, d=18 mm), most spectra show a dominance of methylene (=CH$_2$) over methyl (-CH$_3$) groups. The resulting ratio is shown in Fig. \ref{fig:hc1} plotted against the central position of L1, resulting in a large scatter.  
This is also the case for the ratio of the areas of CH$_{arom.}$ to the sum of the areas of (CH$_{aliph.}$ + CH$_{arom.}$). We note in particular the case for soots produced in the C/O = 1.3 flame, where the aromatic to aliphatic hydrogen ratio is an order of magnitude greater than in most other materials. 

\begin{figure}[htbp]
\begin{center}
\includegraphics[width=90mm]{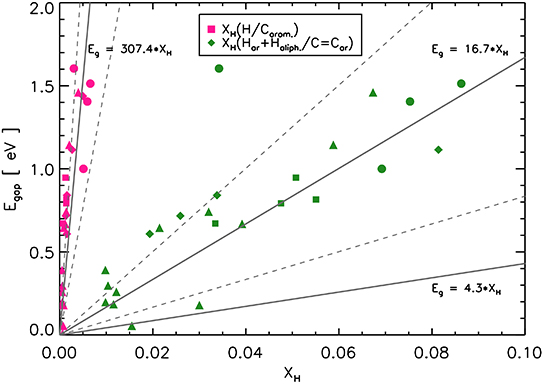}
\caption{Empirically found E$_{g}$ versus X$_H$. We used the aromatic H/C ratio and aromatic+aliphatic H/C ratio, presented as two linear fits for the hydrogen poor (low E$_g$) regime (only soots). When both aromatic and aliphatic CH modes are included, our linear fit approaches the coefficient found for amorphous carbon in \cite{Tamor1990}. The shape of the symbols corresponds to those used in Fig. \ref{fig:hc1}.}
\label{fig:hcxh}
\end{center}
\end{figure}

Carbonaceous materials containing hydrogen can be classified in terms of the hydrogen to carbon ratio \citep{Robertson1987}. To find a representative experimental H/C ratio for our samples, we have considered different combinations of vibrational modes in the infrared spectra. An empirical parameter of the H/C content in our samples is found by taking the ratios of the integrated intensities of the CH$_{arom.}$ mode to that of the aromatic C=C stretching mode ($\sim$ 1600 cm$^{-1}$), (taking A(C=C) $\sim$ 0.175 $\times$ 10$^{-18}$ cm/group \citep{Joblin1994}). We call this the aromatic  H/C ratio.  
Using the empirically found aromatic H/C ratios and the aromatic+aliphatic H/C ratio, that is, N(CH$_{arom.}$ + CH$_{2,3,as}$)/N(C=C), we calculate X$_H$, defined as, 
\begin{equation}
\label{eq:2}
X_H = \frac{H/C}{1+H/C.}
\end{equation}
The result is shown in Fig. \ref{fig:hcxh}. Two linear trends are found for the hydrogen poor (H/C $<<$ 1) carbonaceous samples (soots).

\cite{Tamor1990} provided an empirical linear relationship between the band gap and the H/C ratios of a-C:H materials with E$_g$ = 1.2 - 2 eV, where,

\begin{equation}
\label{tw}
E_g(eV) \simeq 4.3 X_H.
\end{equation}

If we only consider the aromatic H/C ratio, the coefficient is about 100 times larger than the one found by \cite{Tamor1990}. If we include the aliphatic contribution, the fitted slope is still approximately four times larger. Although Eq. \ref{tw} is applicable to a-C:H materials with E$_g$ = 1.2 - 2 eV and our soots have gaps between 0.01 to 1.60 eV, our materials show a linear relationship between the H/C ratio and the gap. 
This demonstrates that not only knowledge of hydrogen bonding but also knowledge of the carbon coordination is essential to understand the physical and chemical structure of carbonaceous materials. For low gap materials, Tamor \& Wu's relationship no longer holds. The evolution of the properties between low gap and high gap materials should be taken into account in optical models of astrophysical carbon.  

\subsection{Carbon structure}

\subsubsection{L$_a$ and N$_{rings}$ versus $\omega_c$(L1)}

The E$_g$ parameter can be used to unveil the structural properties of aromatic absorbers in amorphous carbons \citep{Robertson1987, Robertson1991}, including the average coherence length,  L$_a$ (size of the polyaromatic units), and the number of aromatic rings, N$_R$,  for the largest clusters contributing to the electronic density of states, although not necessarily the most abundant ones \citep{Gadallah2011}. An empirical relation between L$_a$ and E$_g$ was proposed by \cite{Robertson1991}, 
\begin{gather}
L_a ( \text{nm} ) =\Bigg[\frac{0.77}{E_g (eV)} \Bigg]. 
\end{gather}
A similar empirical relation was proposed for a single graphene sheet where the number of six-fold aromatic rings, N$_R$, can be derived from E$_g$ according to the model by \cite{Robertson1987}, in which the gap energy decreases as a function of the size of the graphene layers via the following relation,
\begin{gather}
N_R = \Bigg[\frac{5.8}{E_g (eV)} \Bigg]^2.
\end{gather}

\begin{figure}[htbp]
\begin{center}
\includegraphics[width=90mm]{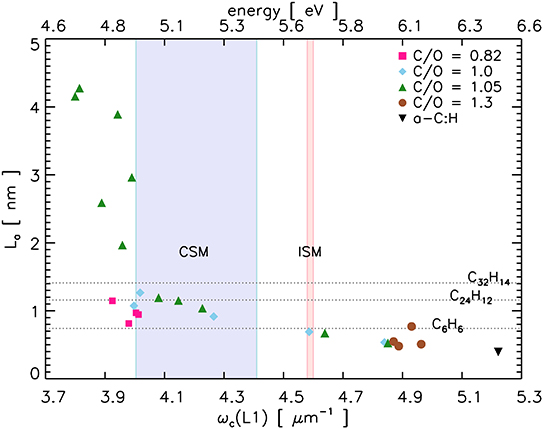}
\caption{Average coherence length, L$_{a}$,  of polyaromatic units of selected laboratory carbons (L$_{a}$$<$ 5 nm) versus $\omega_c$(L1). For comparison, we included the coherence length of Benzene (C$_6$H$_6$), Coronene (C$_{24}$H$_{12}$), and Ovalene (C$_{32}$H$_{14}$) found at \url{http://pah.nist.gov}.}
\label{fig:la}
\end{center}
\end{figure}

\begin{figure}[htbp]
\begin{center}
\includegraphics[width=90mm]{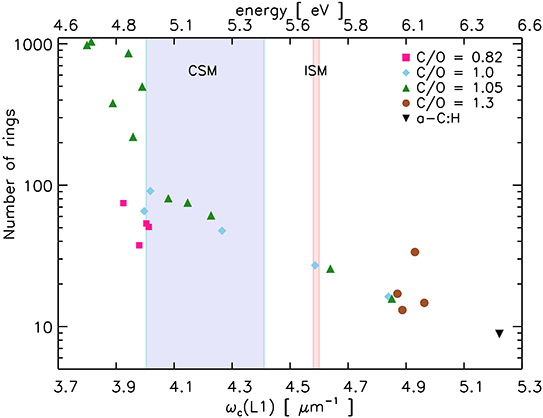}
\caption{Number of rings, N$_{R}$, in the polyaromatic units of our laboratory carbons versus $\omega_c$(L1) suggesting N$_R$ $\sim$60 for the CSM bump and N$_R$ $\sim$25 for the ISM bump.}
\label{fig:nrings}
\end{center}
\end{figure}

We have plotted the E$_g$ derived N$_R$ and L$_a$ values against the position of the main UV bump (also strongly correlated to E$_g$) as shown in Figures \ref{fig:la} and \ref{fig:nrings}. When the position of the L1 peak shifts towards the UV, towards a larger band gap, L$_a$ and  N$_R$ decrease.  For the interstellar UV bump this would correspond to about 25 rings in a polyaromatic unit, and to 65 rings for the circumstellar bump. For the aromatic coherence length, we infer $\sim$0.7 nm for the interstellar bump carrier and $\sim$1 nm for the circumstellar bump carrier. Our soots consist of L$_a$ = 0.5 to 3.3 nm  nanoparticles with corresponding band  gaps of E$_g$ = 0.1 - 1.6 eV. The smallest aromatic length L$_a$ = 0.4 nm was found for the a-C:H with E$_g$ = 2 eV. 

\subsubsection{H/C versus L$_a$ and N$_{rings}$}

We use empirical relationships between the number of rings and the H/C ratio of our materials and compare them to purely aromatic carbons, that is, PAHs. In circum-PAHs, PAH with condensed six-membered rings \citep{Violi2002}, the number of rings grows in a sequence as follows: N$_{rings}$ =1, 7, 19, 37, 61, 91, with  corresponding H/C ratios = 1, 0.5, 0.33, 0.25, 0.2, 0.17. 
To compare the structure of our soots to these PAHs, we will use the relation found in \cite{Jones2012a}  to evaluate the radius of the most compact  aromatic domains, equivalent to L$_a$ for the circum-PAHs,

\begin{equation}
a_R (\text{nm}) =0.09*(2*N_R+\sqrt(N_R)+0.5)^{0.5}.
\end{equation}

\begin{figure}[htbp]
\begin{center}
\includegraphics[width=90mm]{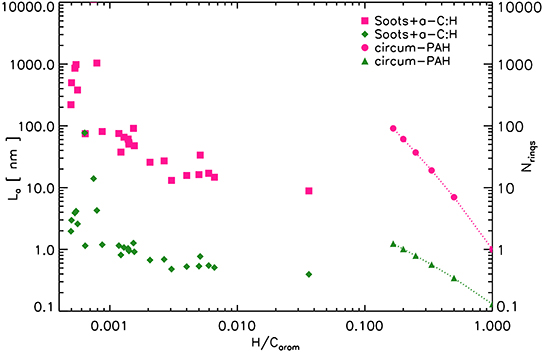}
\caption{Average coherence length, L$_{a}$, and number of rings, N$_{R}$, in the polyaromatic units of our laboratory carbons and circum-PAHs versus the corresponding aromatic H/C ratio.}
\label{fig:hcnrla}
\end{center}
\end{figure}

As show in Fig. \ref{fig:hcnrla} our samples are, as expected, much less hydrogenated than circum-PAHs. For these soots, L$_a$ ranges from 0.3 to 10 nm, while the number of rings in a polyaromatic unit spans from approximately one for the most hydrogenated soot sample to 1000 for the least hydrogenated. Both L$_a$ and N$_R$ decay exponentially as a function of the aromatic H/C ratio following the same trend as PAHs. Similarly, to find the number of carbon atoms in these polyaromatic units, we use the following equation from \citep{Jones2012a},

\begin{equation}
\label{eq:nc}
N_C=2*N_R+3.5*\sqrt(N_R)+0.5.
\end{equation}

\begin{figure}[htbp]
\begin{center}
\includegraphics[width=84mm]{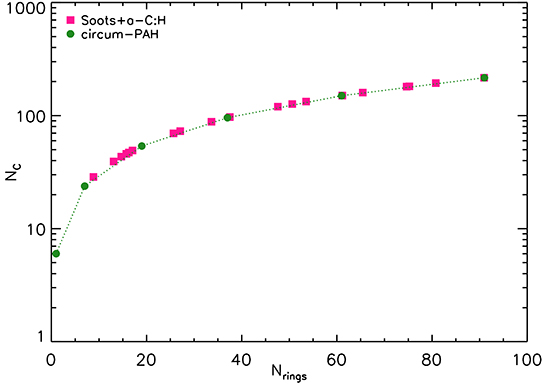}
\caption{Number of carbon atoms (N$_C$) versus the number of rings N$_{R}$ in the polyaromatic units of the laboratory samples presented here (soots and a-C:H) and circum-PAHs,  using Eq. \ref{eq:nc}.}
\label{fig:arnc}
\end{center}
\end{figure}

As shown in Fig. \ref{fig:arnc}, the polyaromatic units in our soot samples contain between 20 and 100 carbon atoms.
These are in the same range as those discussed in \cite{Steglich2010}, who proposed, in accordance with semi-empirical calculations, that a distribution of large PAH molecules with a mean size of 50-60 carbon atoms can produce a bump centered at $\sim$217.5 nm. 

\section{Discussion}
\label{sec:dis}

\begin{figure}[htbp]
\begin{center}
\includegraphics[width=90mm]{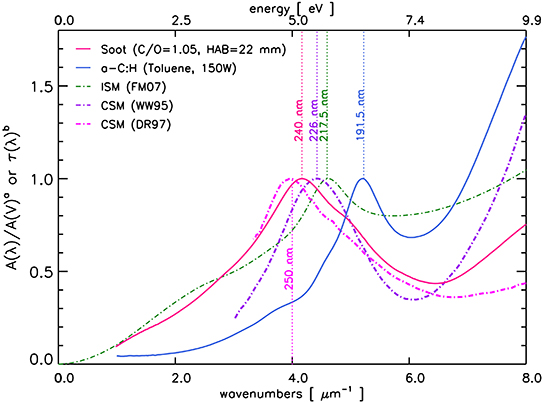}
\caption{Normalized laboratory soot and a-C:H visible-VUV spectra compared to the average ISM extinction curve  \citep{Fitzpatrick2007} (FM07), and the CSM extinction curves of the circumstellar envelope of HD 213985 \citep{Waelkens1995} (WW95) and V348 Sagittarii \citep{Drilling1997} (DR97), for which the original spectrum has been heavily filtered. All spectra are re-normalized to the maximum intensity of their $\pi$-$\pi^*$ resonance.}
\label{fig:ismcsm}
\end{center}
\end{figure}

The transmission measurements from the VUV to the mid-infrared of several carbonaceous samples have allowed us to perform a large spectroscopic study covering the main electronic and vibrational transitions. We find the following correlations: 
\begin{itemize}
\item The position of the UV L1 feature ($\pi$-$\pi^*$) is correlated to the position of the sp$^3$ carbon defect, and to the position of the aromatic C=C stretching mode. This reveals a relation between electronic and vibrational band positions in soots.  
\item The Tauc optical gap, E$_g$, is correlated to the position of the $\pi$-$\pi^*$ band, $\omega_c$(L1). Both E$_g$ and $\omega_c$(L1) are relevant parameters allowing the classification of both soots and a-C:Hs. 
\item The H/C ratio of these carbonaceous samples is correlated to E$_g$ and therefore to $\omega_c$(L1). For soots, the empirical relationship between the H/C ratio and the gap differs from that found for highly hydrogenated carbonaceous materials.  
\end{itemize}

The wide variety of our laboratory carbonaceous dust allows us to explore the spectral properties of the observed ISM and CSM extinction curves (e.g., Fig \ref{fig:ismcsm}), from the VUV to the IR:
\begin{itemize} 
\item Spectroscopy in the VUV (120 $<$ $\lambda$ $<$ 200 nm) has allowed us to identify an absorption band  in the VUV present in all our samples: the G3 band, with a fitted position centered at $\sim$9 $\mu$m$^{-1}$ ($\sim$110 nm) and an integrated optical depth ratio of approximately two with respect to L1. G3 is attributed to  $\sigma$-$\sigma^*$ transitions and may contribute to the FUV rise identified in \cite{Fitzpatrick2007}. 
\item For the average ISM bump, centered at 217.5 nm, we infer the following spectral properties of the carrier: E$_g$ $\sim$1.2 eV,  $\omega_c$(G3) $\sim$9.5$\mu$m$^{-1}$, $\omega_c$(sp$^3$ defect) $\sim$1280 cm$^{-1}$, $\omega_c$(C=C) $\sim$1600 cm$^{-1}$.
\item For the average CSM bump, centered $\sim$226-250 nm, we infer the following spectral properties of the carrier: E$_g$ $\sim$0.7 eV, $\omega_c$(G3) $\sim$9 $\mu$m$^{-1}$, $\omega_c$(sp$^3$ defect) $\sim$1250 cm$^{-1}$, $\omega_c$(C=C) $\sim$1592 cm$^{-1}$. 
\end{itemize}

The connection between ultraviolet extinction data and infrared emission has been possible thanks to early datasets from the IUE and the Infrared Astronomical Satellite (IRAS), observing towards the same lines of sight. \cite{Cox1987} showed the absence of correlation between the mid-infrared excess and the normalized intensity of the UV bump, but later work by \cite{Cardelli1989} showed that their bump normalization was not appropriate. An extended study by \cite{Jenniskens1993}  and later on by \cite{Boulanger1994} revealed a link between mid-infrared emission bands (AIBs) and the UV bump absorption. In contrast, they found no relation between the AIBs and the FUV rise. 

Correlations between the positions of UV and infrared absorption bands are found for the carbonaceous samples presented in this paper. These spectral parameters suggest that the carrier of the ISM bump is likely found at the frontier of class B/A of the AIBs, while the carrier of the CSM bump is found at the frontier of class B/C. Based on the Tauc optical gap and using available empirical formulae for carbons, the structure of the ISM UV bump carrier implies an average polyaromatic unit size $\sim$0.7 nm and for the CSM UV bump carrier, $\sim$1.1 nm, within nanoparticles. 

\begin{figure}[htbp!]
\begin{center}
\includegraphics[width=90mm]{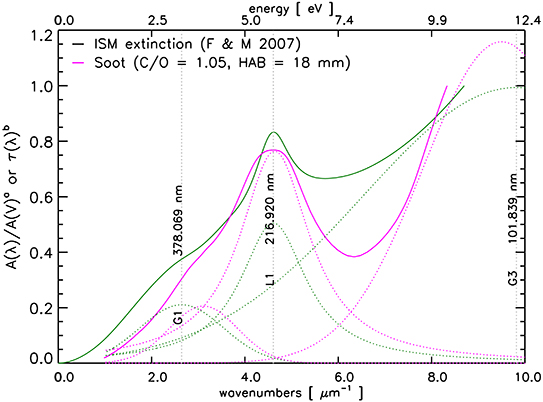}
\caption{Deconvolution of the average interstellar extinction curve \citep{Fitzpatrick2007} ($^{a}$normalized to its maximum) following the deconvolution procedure of laboratory soots described in Sect. \ref{sec:uvfit}. The L1 peak is centered at $\omega_c$ = 4.609 $\mu$m$^{-1}$ with a FWHM = 1.779 $\mu$m$^{-1}$. While the position is unchanged with respect to the parametrization of \cite{Fitzpatrick2007}, the original FWHM value is almost doubled. For comparison, the spectra of a laboratory soot prepared in a flame with C/O = 1.05 and HAB = 18 mm is included, showing a similarly intense and wide $\pi$-$\pi^*$ transition, a G1 peak, and a slightly less intense FUV rise.}
\label{fig:ISM}
\end{center}
\end{figure}

While the position of the interstellar 217.5 nm bump (4.592 $\mu$m$^{-1}$) is stable, its FWHM varies \citep{Fitzpatrick2005}. The FWHM of the $\pi$-$\pi^*$ band of our carbonaceous analogs, $\gamma$(L1), is correlated to the position $\omega_c$(L1). None of our samples fulfill simultaneously the observed constraints on the UV bump (position, FWHM) in interstellar spectra \citep{Fitzpatrick2007}. In Fig. \ref{fig:ISM}, we take the average interstellar extinction curve and perform a deconvolution as done for our laboratory samples to explore the effect of the deconvolution on the same basis. A direct comparison of the FWHM of a laboratory soot (1.88 $\mu$m$^{-1}$) to the newly parametrized FWHM (1.78 $\mu$m$^{-1}$) from the average interstellar extinction curve indicates that laboratory polyaromatic solid carbons can approximate the average carrier of the UV bump. However, it is more likely than more than one kind of carbon is measured towards a line of sight, and that more hydrogenated carbons contribute strongly to the FUV rise as shown in \cite{Gavilan2016a} and in Fig. \ref{fig:ismcsm}, where our laboratory soot and a-C:H spectra are compared to extinction curves in the ISM and CSM. 

\section{Conclusions}
\label{sec:5}

We have produced laboratory analogs to interstellar carbonaceous dust characterized by their wide degree of hydrogenation and differences in polyaromatic structure. This influences the spectral signatures, in particular the position of the UV bump, attributed to $\pi$-$\pi^*$ electronic transitions. Our carbonaceous analogs consist of soots, characterized by low H/C and E$_g$ values. A hydrogenated a-C:H with E$_g$ $\sim$2 eV was also produced. 
The UV to mid-infrared spectroscopy has allowed us to find correlations between the electronic transitions  and vibrational modes and to deduce structural properties such as the average length ($\sim$0.7 - 1.1 nm) of polyaromatic units within nanoparticles.

These laboratory correlations suggest that for the interstellar and circumstellar UV bump carriers, infrared signatures are expected in classes B/A and B/C of the AIBs respectively. More hydrogenated carbons contribute to the FUV rise in agreement with its weak correlation with the UV bump. Observations are then not only sensitive to local environmental conditions but also to the transition of dust from circum- to interstellar environments. As shown by our measurements, disordered polyaromatic carbonaceous grains are viable carriers of the observed UV interstellar and circumstellar extinction. \\

\vspace{0.01cm}
{ \footnotesize
\textit{Acknowledgments.} VUV measurements were performed at the DISCO beamline of the SOLEIL synchrotron (projects: 20141074 \& 20130778). This work has been supported by the French program PCMI, and the ANR COSMISME project (grant ANR-2010-BLAN-0502). K.C.L. thanks the Vietnamese government for a doctoral scholarship. L.G. thanks the Centre National d'\'Etudes Spatiales (CNES) for a post-doctoral fellowship.} \\

\bibliography{vuv}

\begin{center}
Online Materials
\end{center}

\begin{appendix}
\section{}
\label{app1}
Here we provide the additional Figures and Tables that are referenced in the main body of the text. These are organized as VUV spectra (1-10 $\mu$m$^{-1}$), and the mid-infrared spectra of the 3.4 $\mu$m band region and the 5-20 $\mu$m band region.

\begin{figure*}
\centering
\label{fig:a1}
\caption{UV-VUV and mid-infrared spectral deconvolution for soot samples prepared with a C/O = 0.82 combustion flame, retrieved at a HAB of \textbf{(a)} 20 mm to \textbf{(g)} 30 mm. }
\vspace{1em}
\subfloat[ 20 mm ]{\includegraphics[width = 54mm]{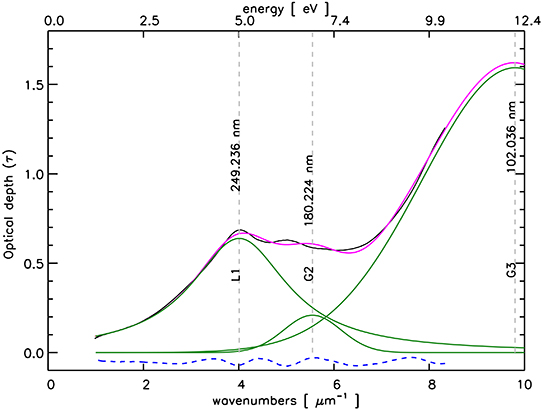}} \hspace{0.5cm}
\subfloat[ 20 mm ]{\includegraphics[width = 54mm]{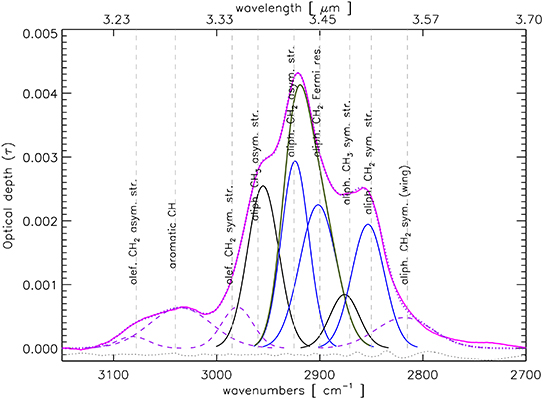}} \hspace{0.5cm}
\subfloat[ 20 mm ]{\includegraphics[width = 54mm]{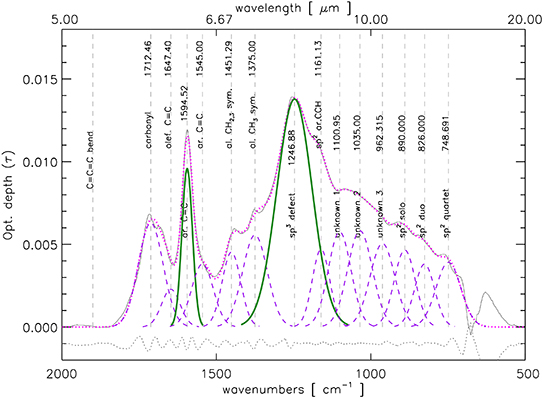}}\\ \vspace{-1em}
\subfloat[ 21 mm ]{\includegraphics[width = 54mm]{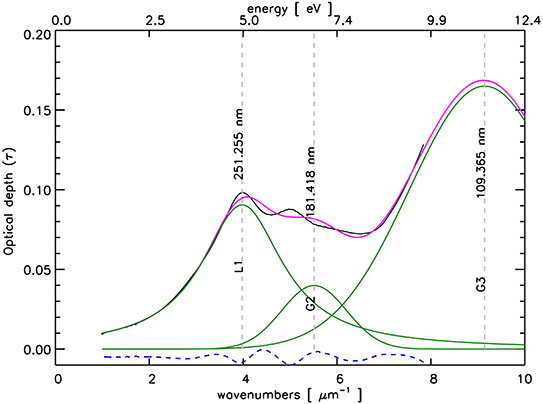}} \hspace{0.5cm}
\subfloat[ 21 mm ]{\includegraphics[width = 54mm]{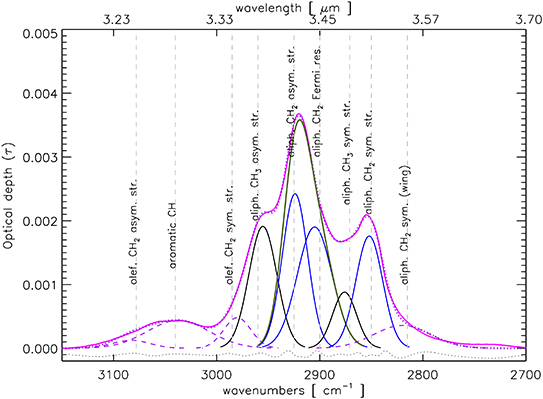}} \hspace{0.5cm}
\subfloat[ 21 mm ]{\includegraphics[width = 54mm]{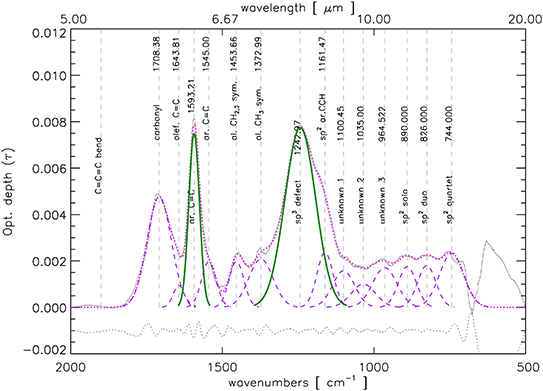}}\\ \vspace{-1em}
\subfloat[ 22 mm ]{\includegraphics[width = 54mm]{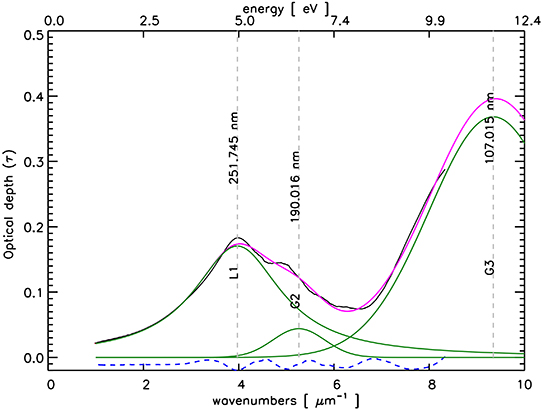}} \hspace{0.5cm}
\subfloat[ 22 mm ]{\includegraphics[width = 54mm]{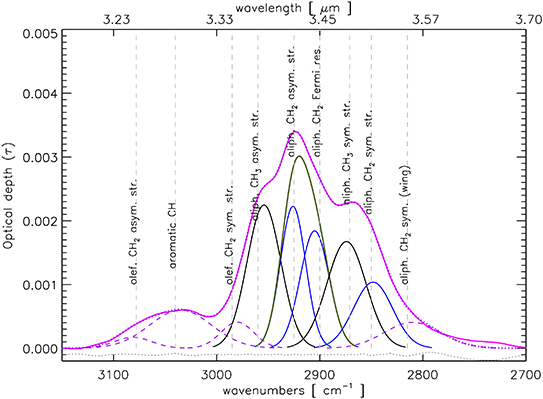}} \hspace{0.5cm}
\subfloat[ 22 mm ]{\includegraphics[width = 54mm]{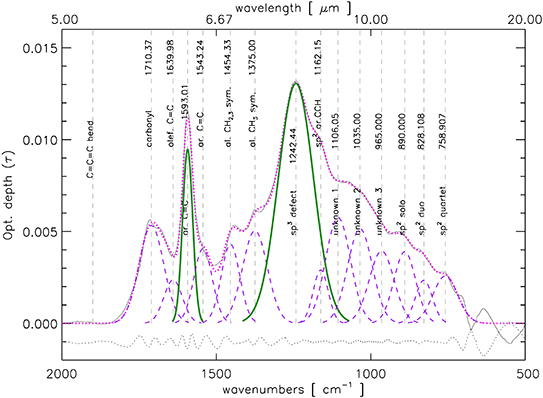}}\\ \vspace{-1em}
\subfloat[ 30 mm ]{\includegraphics[width = 54mm]{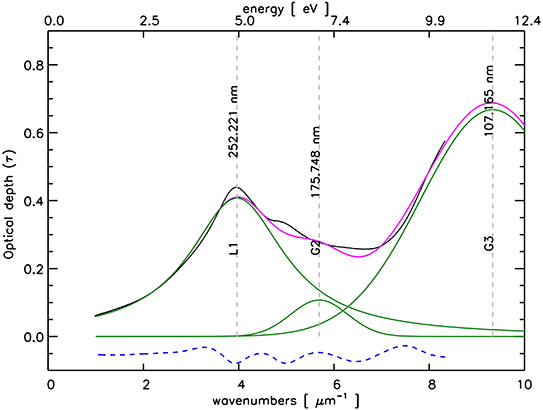}} \hspace{0.5cm}
\subfloat[ 30 mm ]{\includegraphics[width = 54mm]{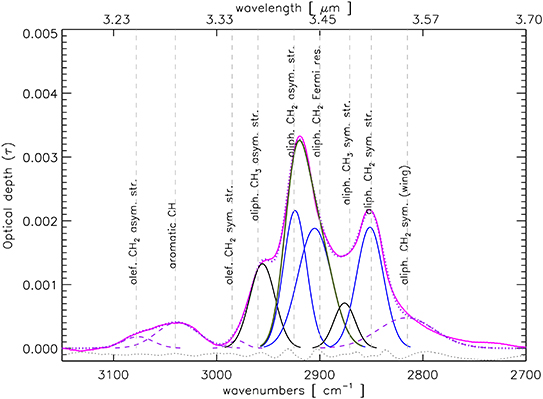}} \hspace{0.5cm}
\subfloat[ 30 mm ]{\includegraphics[width = 54mm]{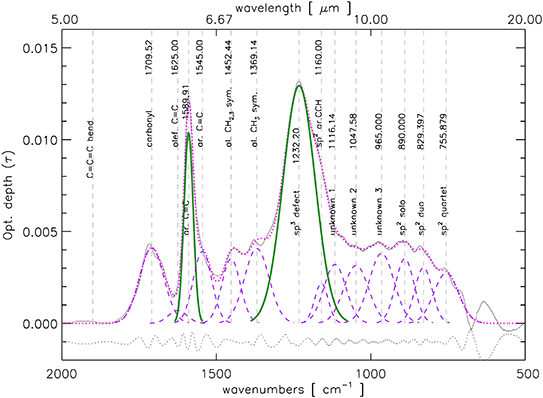}}
\end{figure*}

\begin{figure*}
\centering
\label{fig:a2}
\caption{UV-VUV and mid-infrared spectral deconvolution for soot samples prepared with a C/O = 1 combustion flame, retrieved at a HAB of \textbf{(a)} 18 mm to \textbf{(i)} 39 mm.}
\vspace{1em}
\subfloat[ 18 mm ]{\includegraphics[width = 54mm]{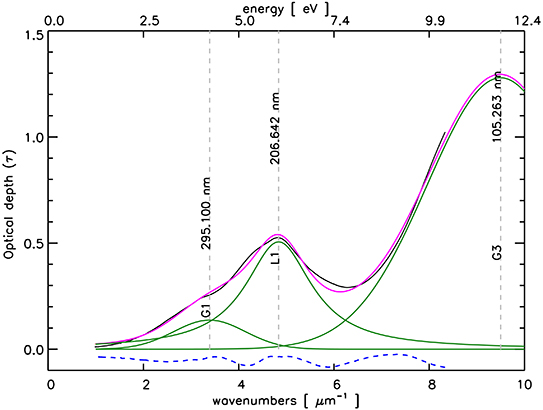}} \hspace{0.5cm}
\subfloat[ 18 mm ]{\includegraphics[width = 54mm]{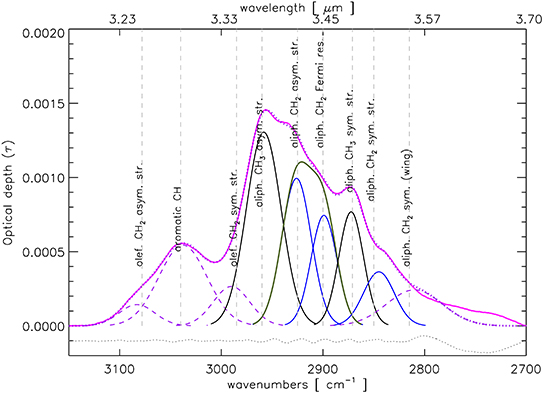}} \hspace{0.5cm}
\subfloat[ 18 mm ]{\includegraphics[width = 54mm]{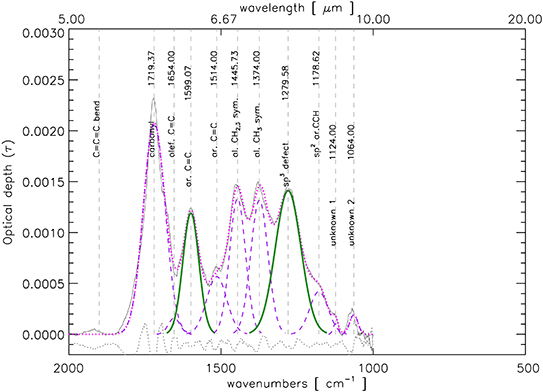}}\\ \vspace{-1em}
\subfloat[ 21 mm ]{\includegraphics[width = 54mm]{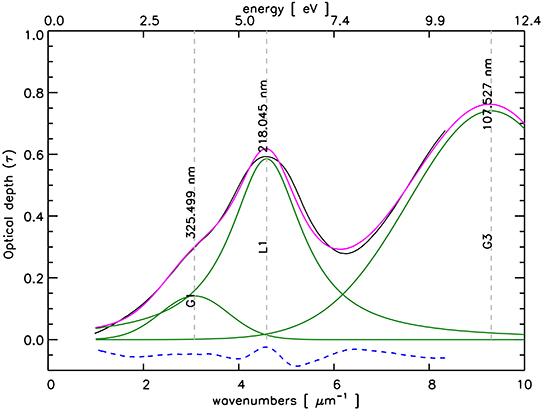}} \hspace{0.5cm}
\subfloat[ 21 mm ]{\includegraphics[width = 54mm]{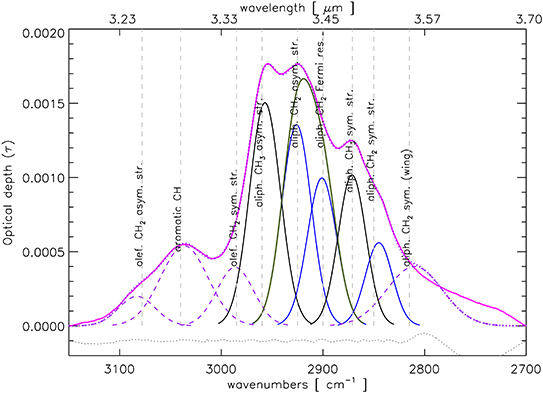}} \hspace{0.5cm}
\subfloat[ 21 mm ]{\includegraphics[width = 54mm]{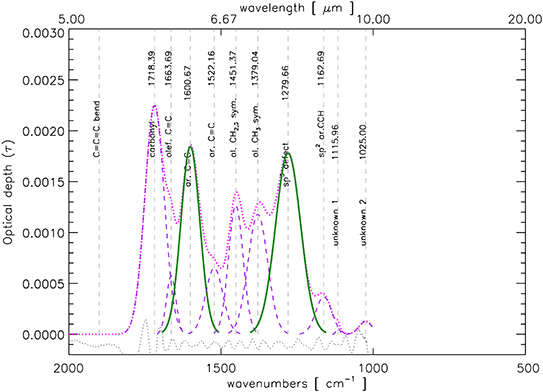}}\\ \vspace{-1em}
\subfloat[ 25 mm ]{\includegraphics[width = 54mm]{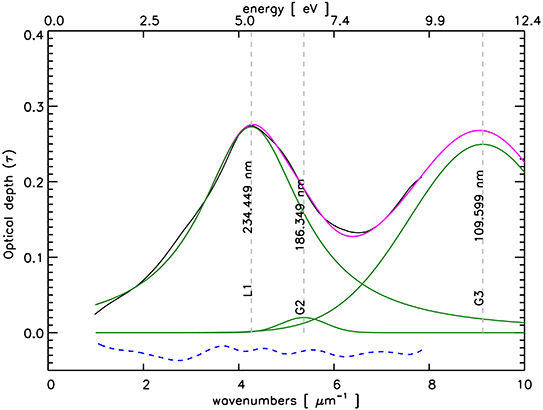}} \hspace{0.5cm}
\subfloat[ 25 mm ]{\includegraphics[width = 54mm]{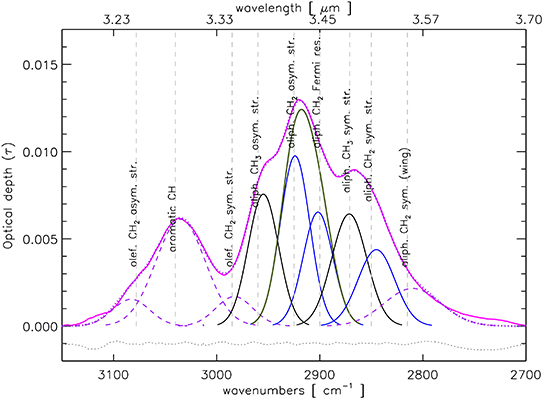}} \hspace{0.5cm}
\subfloat[ 25 mm ]{\includegraphics[width = 54mm]{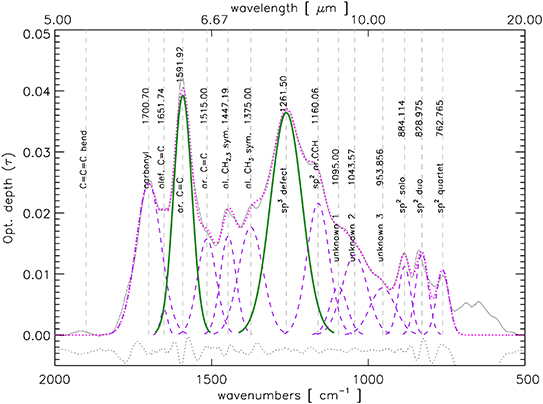}}\\ \vspace{-1em}
\subfloat[ 30 mm ]{\includegraphics[width = 54mm]{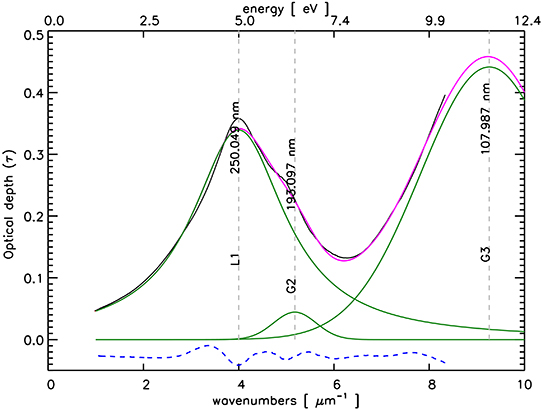}} \hspace{0.5cm}
\subfloat[ 30 mm ]{\includegraphics[width = 54mm]{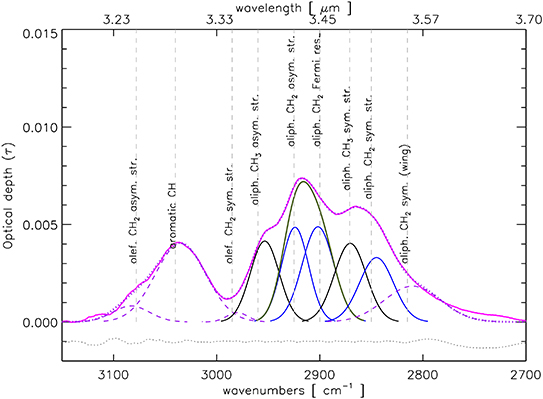}} \hspace{0.5cm}
\subfloat[ 30 mm ]{\includegraphics[width = 54mm]{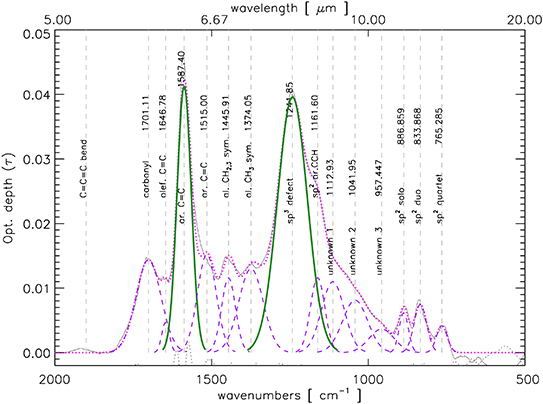}}\\ \vspace{-1em}
\subfloat[ 39 mm ]{\includegraphics[width = 54mm]{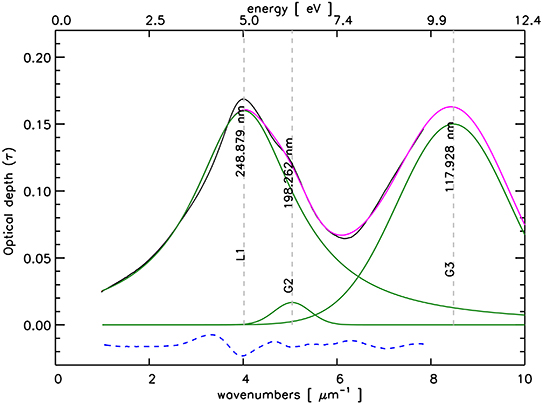}} \hspace{0.5cm}
\subfloat[ 39 mm ]{\includegraphics[width = 54mm]{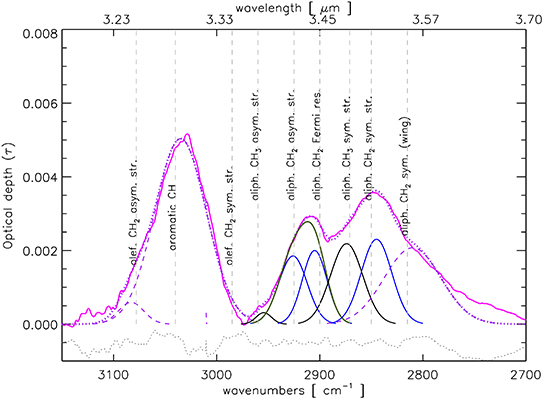}} \hspace{0.5cm}
\subfloat[ 39 mm ]{\includegraphics[width = 54mm]{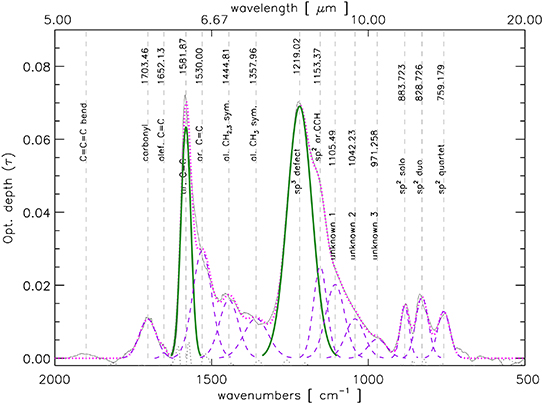}}
\end{figure*}

\newpage

\begin{figure*}
\centering
\label{fig:a3}
\caption{UV-VUV and mid-infrared spectral deconvolution for soot samples prepared with a C/O = 1.05 combustion flame, retrieved at a HAB of \textbf{(a)} 18 mm to \textbf{(e)} 20 mm.}
\vspace{1em}
\subfloat[ 14 mm ]{\includegraphics[width = 54mm]{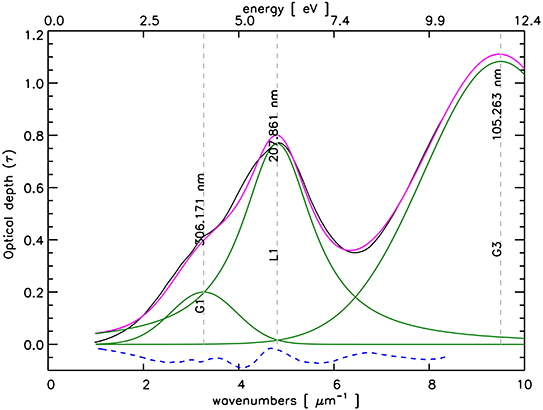}} \hspace{0.5cm}
\subfloat[ 14 mm ]{\includegraphics[width = 54mm]{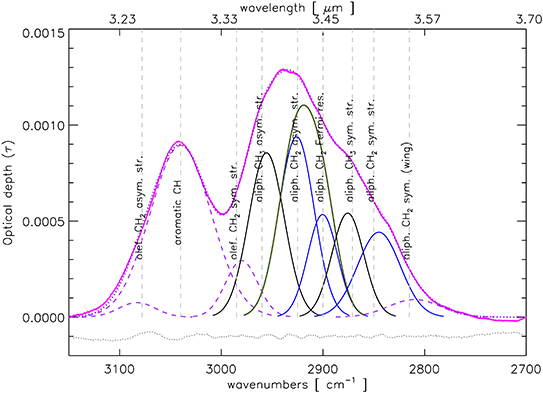}} \hspace{0.5cm}
\subfloat[ 14 mm ]{\includegraphics[width = 54mm]{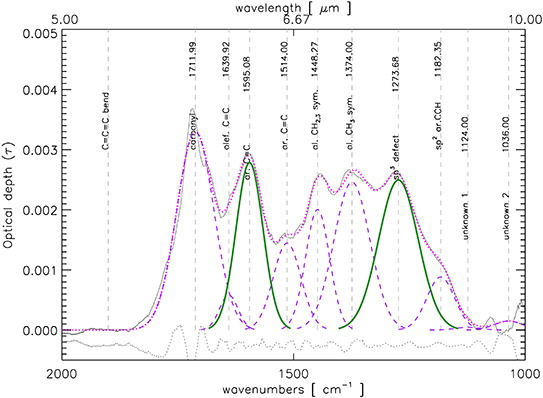}}\\ \vspace{-1em}
\subfloat[ 18 mm ]{\includegraphics[width = 54mm]{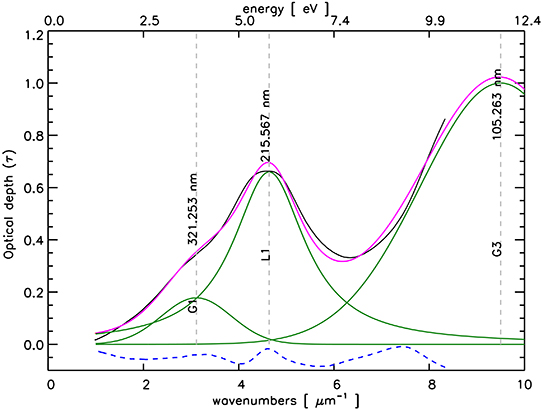}} \hspace{0.5cm}
\subfloat[ 18 mm ]{\includegraphics[width = 54mm]{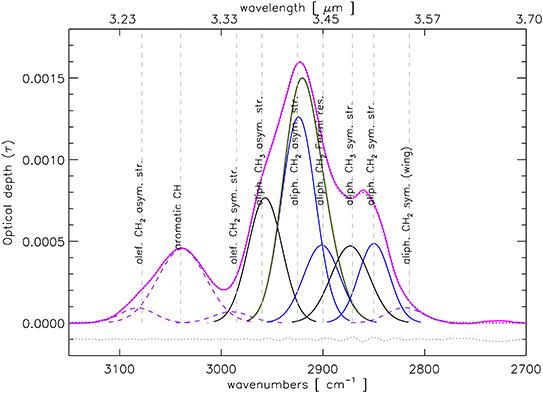}} \hspace{0.5cm}
\subfloat[ 18 mm ]{\includegraphics[width = 54mm]{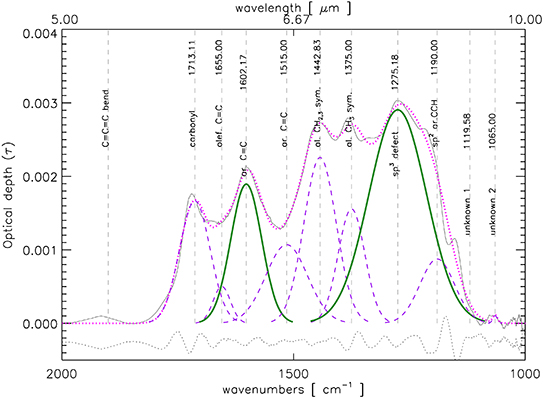}}\\ \vspace*{-1em}
\subfloat[ 22 mm ]{\includegraphics[width = 54mm]{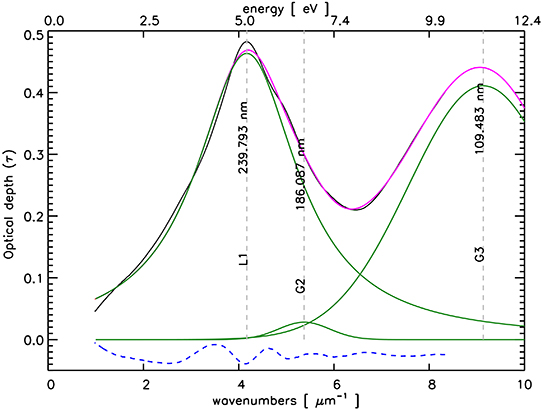}} \hspace{0.5cm}
\subfloat[ 22 mm ]{\includegraphics[width = 54mm]{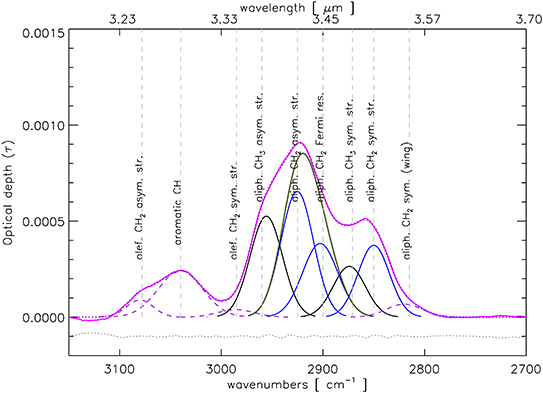}} \hspace{0.5cm}
\subfloat[ 22 mm ]{\includegraphics[width = 54mm]{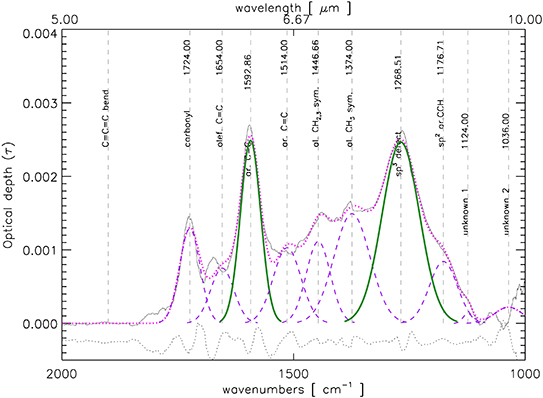}}
\end{figure*}

\newpage

\begin{figure*}
\centering
\label{fig:a4}
\caption{UV-VUV and mid-infrared spectral deconvolution for soot samples prepared with a C/O = 1.05 combustion flame, retrieved at a HAB of \textbf{(a)} 22 mm to \textbf{(j)} 30 mm. }
\vspace{1em}
\subfloat[ 22 mm ]{\includegraphics[width = 54mm]{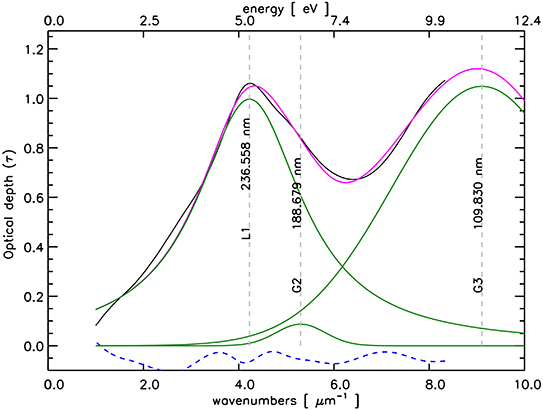}} \hspace{0.5cm}
\subfloat[ 22 mm ]{\includegraphics[width = 54mm]{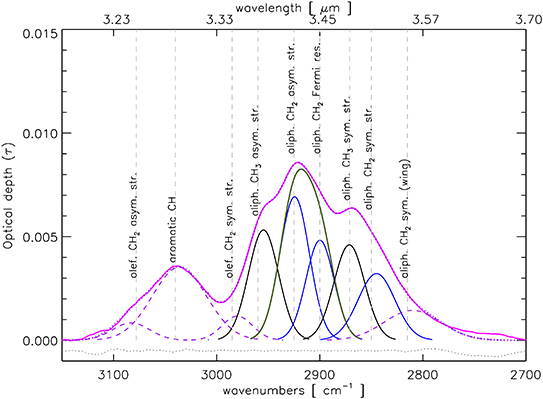}} \hspace{0.5cm}
\subfloat[ 22 mm ]{\includegraphics[width = 54mm]{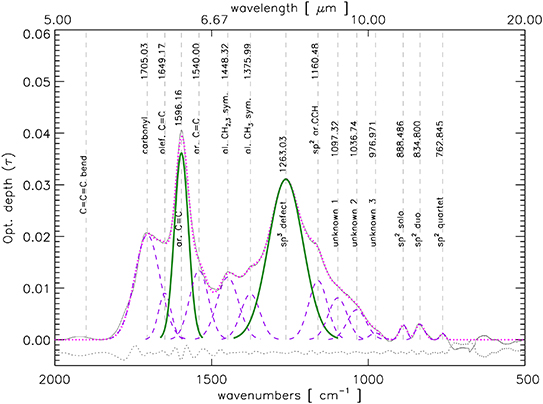}}\\ \vspace{-1em}
\subfloat[ 24 mm ]{\includegraphics[width = 54mm]{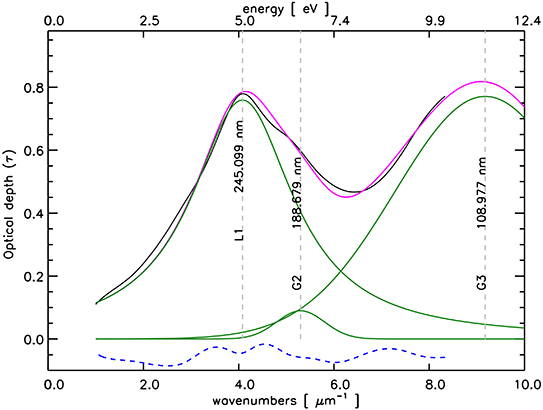}}\hspace{0.5cm}
\subfloat[ 24 mm ]{\includegraphics[width = 54mm]{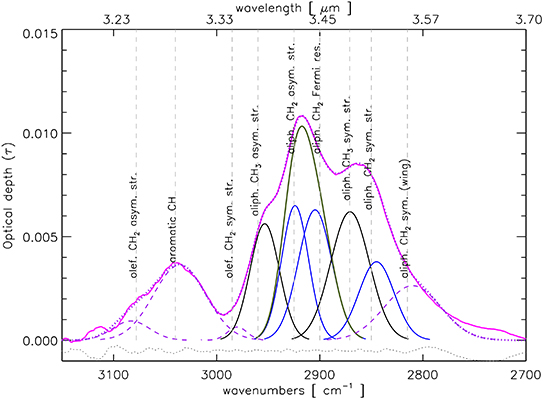}} \hspace{0.5cm}
\subfloat[ 24 mm ]{\includegraphics[width = 54mm]{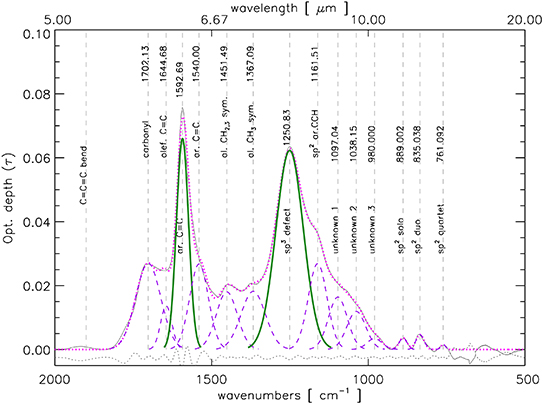}}\\ \vspace*{-1em}
\subfloat[ 26 mm ]{\includegraphics[width = 54mm]{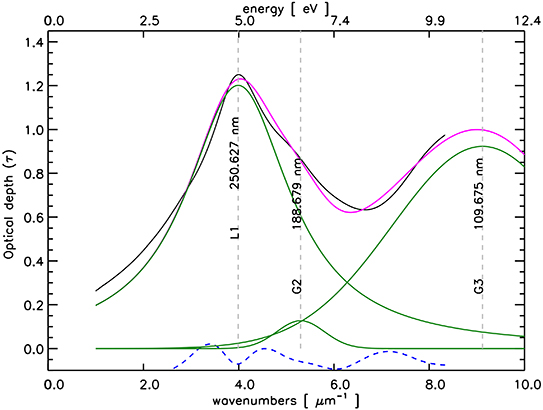}}\hspace{0.5cm}
\subfloat[ 26 mm ]{\includegraphics[width = 54mm]{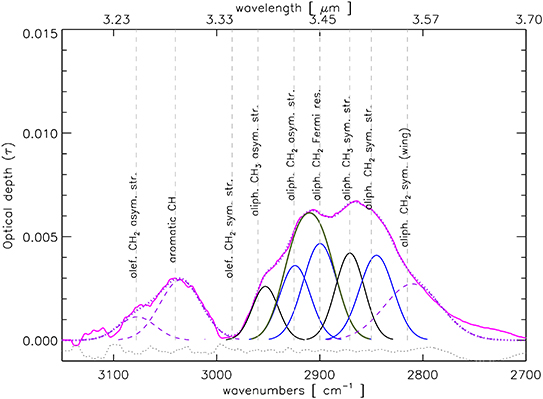}} \hspace{0.5cm}
\subfloat[ 26 mm ]{\includegraphics[width = 54mm]{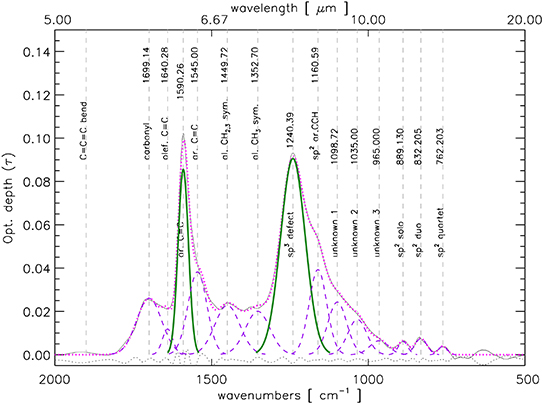}}\\ \vspace*{-1em}
\subfloat[ 28 mm ]{\includegraphics[width = 54mm]{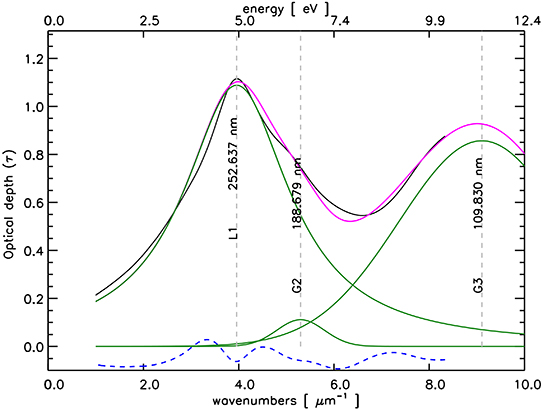}}\hspace{0.5cm}
\subfloat[ 28 mm ]{\includegraphics[width = 54mm]{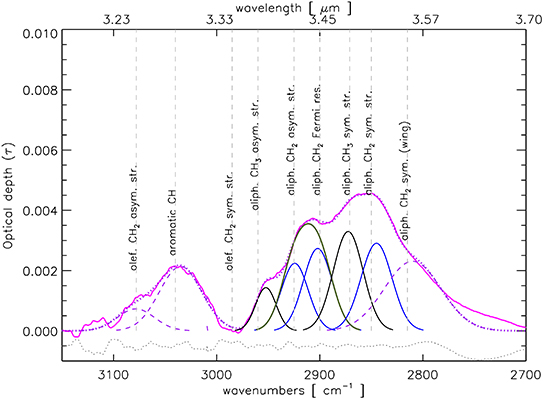}} \hspace{0.5cm}
\subfloat[ 28 mm ]{\includegraphics[width = 54mm]{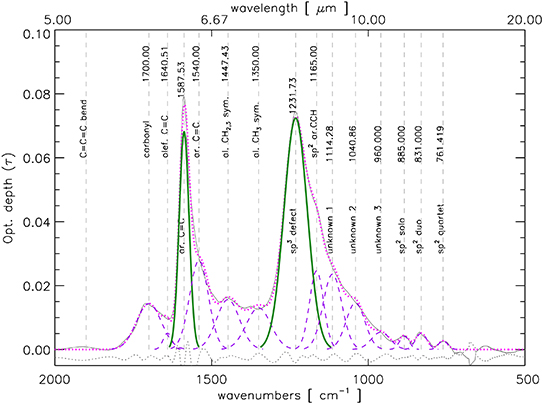}}\\ \vspace*{-1em}
\subfloat[ 30 mm ]{\includegraphics[width = 54mm]{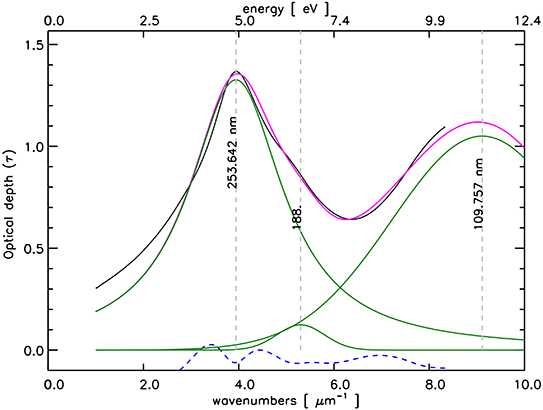}}\hspace{0.5cm}
\subfloat[ 30 mm ]{\includegraphics[width = 54mm]{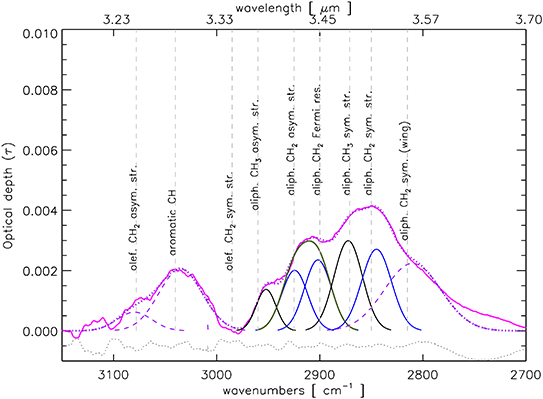}} \hspace{0.5cm}
\subfloat[ 30 mm ]{\includegraphics[width = 54mm]{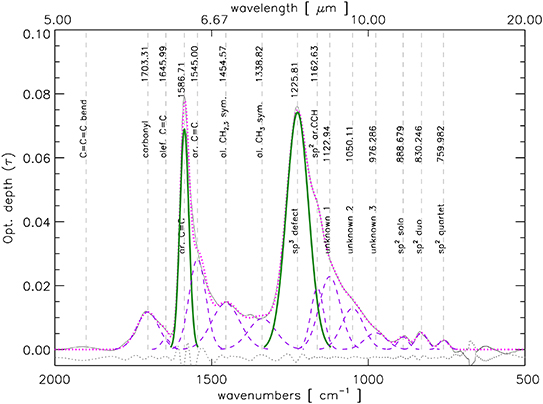}}  
\end{figure*}

\newpage

\begin{figure*}
\centering
\label{fig:a5}
\caption{UV-VUV and mid-infrared spectral deconvolution for soot samples prepared with a C/O = 1.05 combustion flame, retrieved at a HAB of \textbf{(a)} 32 mm to \textbf{(j)} 50 mm. }
\vspace{1em}
\subfloat[ 32 mm ]{\includegraphics[width = 54mm]{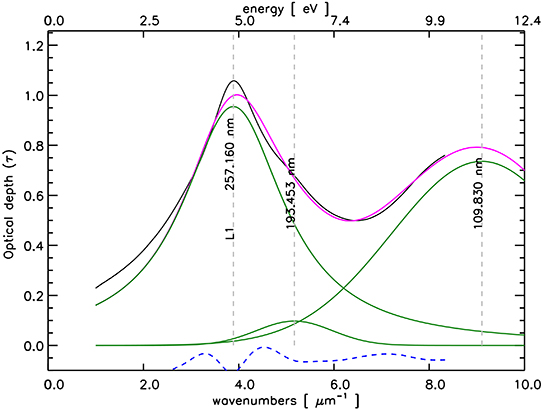}} \hspace{0.5cm}
\subfloat[ 32 mm ]{\includegraphics[width = 54mm]{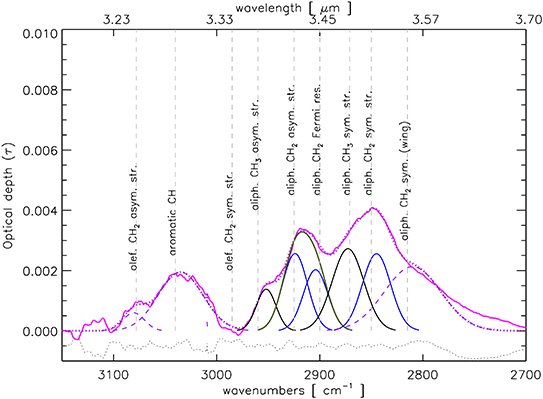}} \hspace{0.5cm}
\subfloat[ 32 mm ]{\includegraphics[width = 54mm]{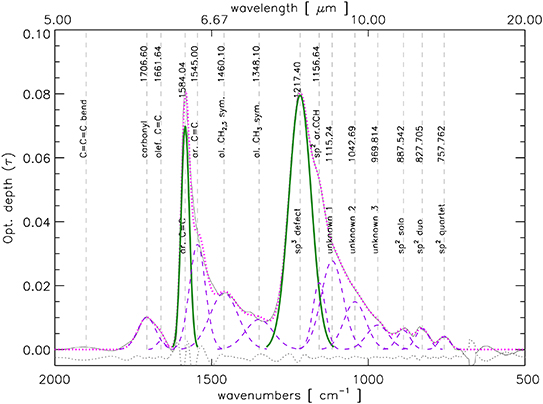}}\\ \vspace{-1em}
\subfloat[ 34 mm ]{\includegraphics[width = 54mm]{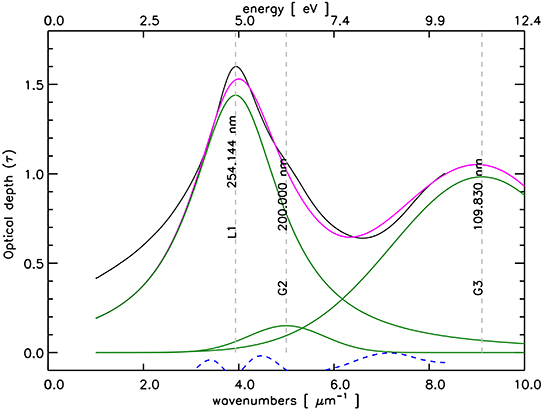}}\hspace{0.5cm}
\subfloat[ 34 mm ]{\includegraphics[width = 54mm]{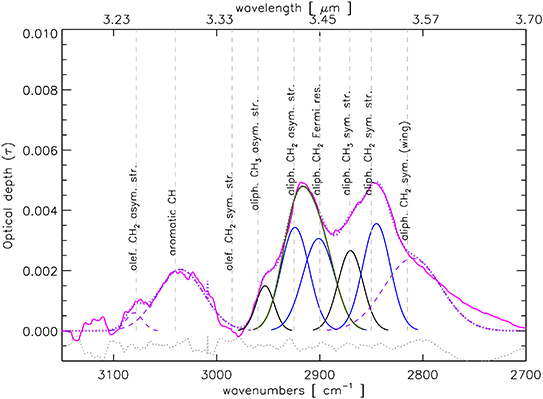}} \hspace{0.5cm}
\subfloat[ 34 mm ]{\includegraphics[width = 54mm]{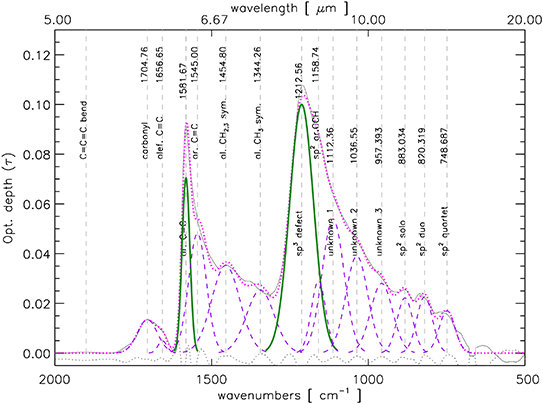}}\\ \vspace*{-1em}
\subfloat[ 36 mm ]{\includegraphics[width = 54mm]{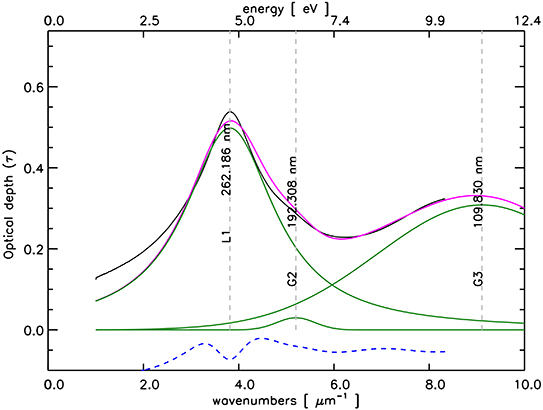}}\hspace{0.5cm}
\subfloat[ 36 mm ]{\includegraphics[width = 54mm]{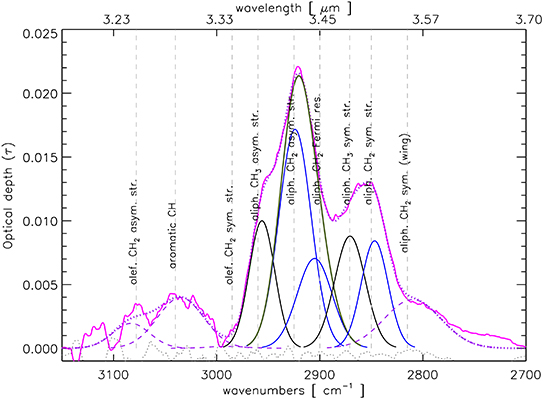}} \hspace{0.5cm}
\subfloat[ 36 mm ]{\includegraphics[width = 54mm]{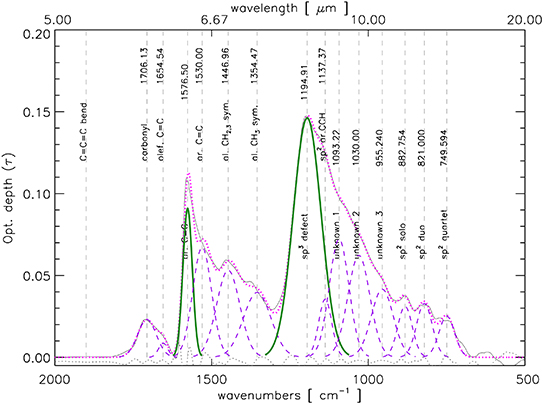}}\\ \vspace*{-1em}
\subfloat[ 42 mm ]{\includegraphics[width = 54mm]{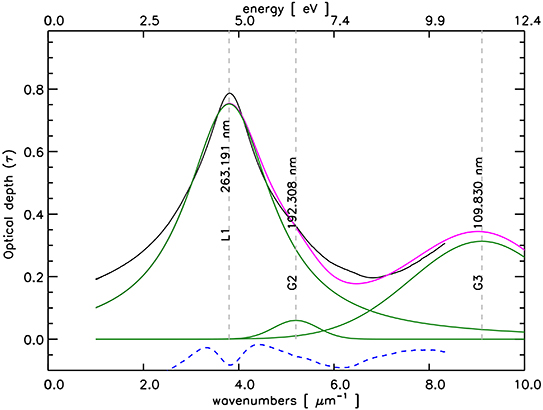}}\hspace{0.5cm}
\subfloat[ 42 mm ]{\includegraphics[width = 54mm]{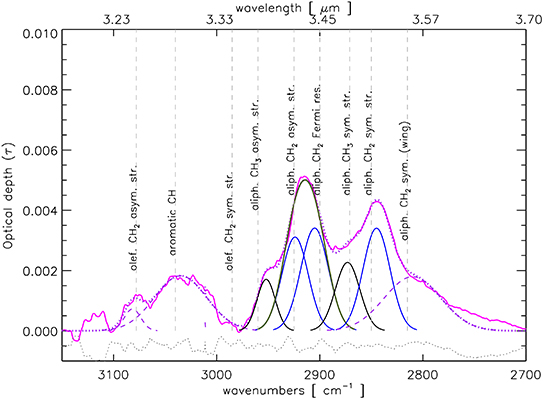}} \hspace{0.5cm}
\subfloat[ 42 mm ]{\includegraphics[width = 54mm]{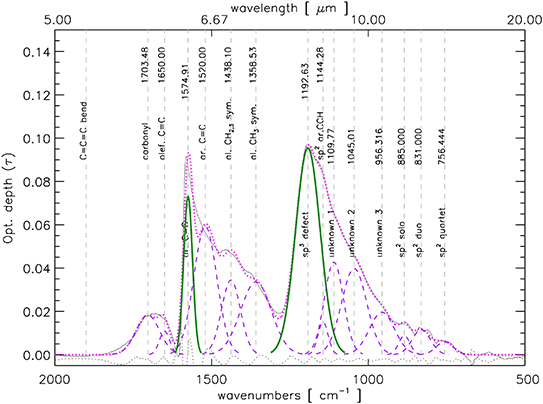}}\\ \vspace*{-1em}
\subfloat[ 50 mm ]{\includegraphics[width = 54mm]{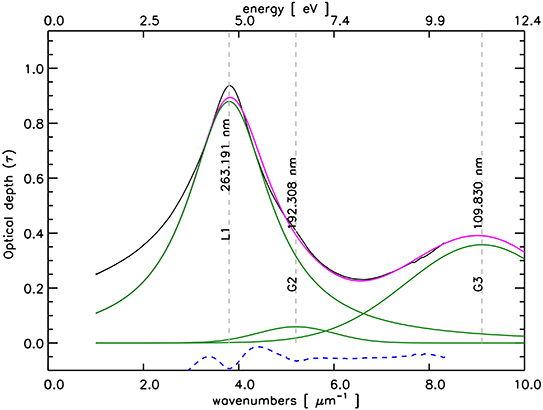}}\hspace{0.5cm}
\subfloat[ 50 mm ]{\includegraphics[width = 54mm]{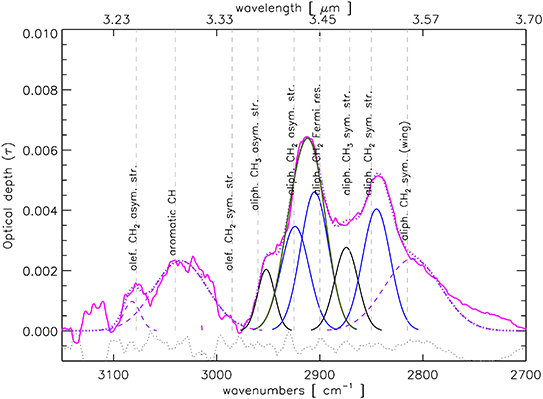}} \hspace{0.5cm}
\subfloat[ 50 mm ]{\includegraphics[width = 54mm]{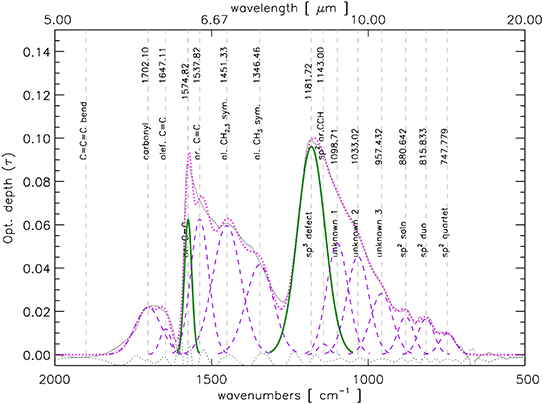}}  
\end{figure*}

\newpage

\begin{figure*}
\centering
\label{fig:a6}
\caption{UV-VUV and mid-infrared spectral deconvolution for soot samples prepared with a C/O = 1.3 combustion flame, retrieved at a HAB of \textbf{(a)} 25 mm to \textbf{(g)} 39 mm. }
\vspace{1em}
\subfloat[ 25 mm ]{\includegraphics[width = 54mm]{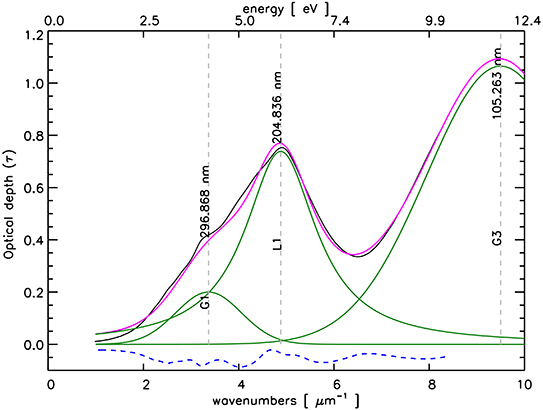}} \hspace{0.5cm}
\subfloat[ 25 mm ]{\includegraphics[width = 54mm]{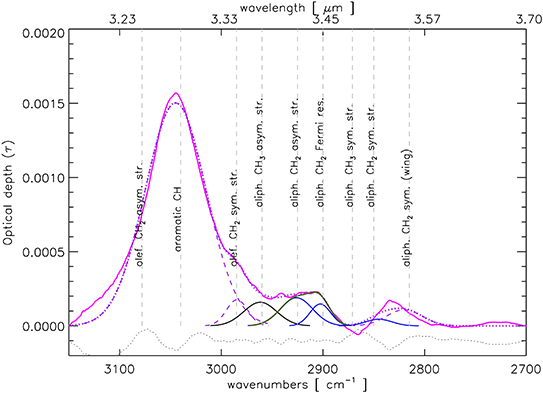}} \hspace{0.5cm}
\subfloat[ 25 mm ]{\includegraphics[width = 54mm]{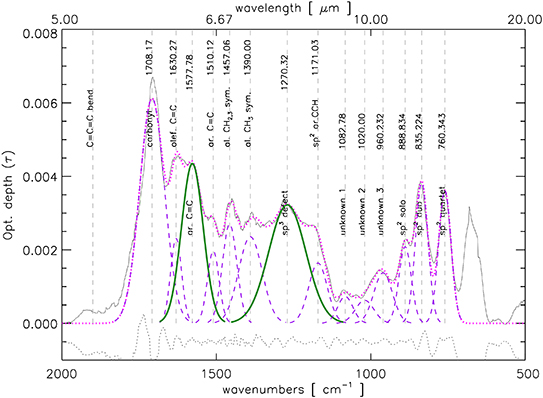}}\\ \vspace{-1em}
\subfloat[ 27 mm ]{\includegraphics[width = 54mm]{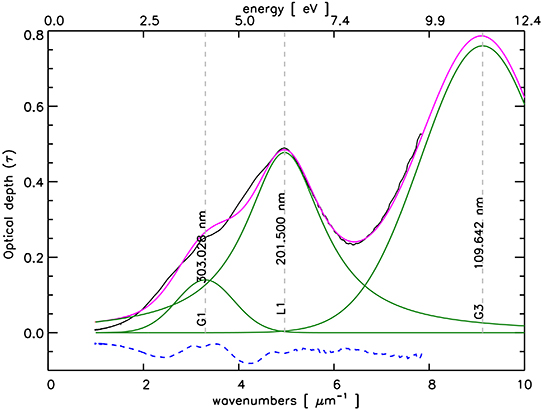}} \hspace{0.5cm}
\subfloat[ 27 mm ]{\includegraphics[width = 54mm]{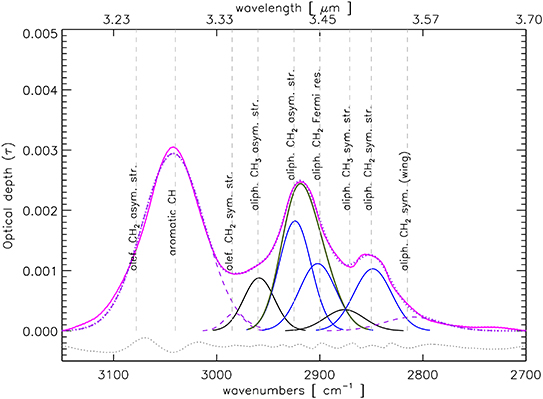}} \hspace{0.5cm}
\subfloat[ 27 mm ]{\includegraphics[width = 54mm]{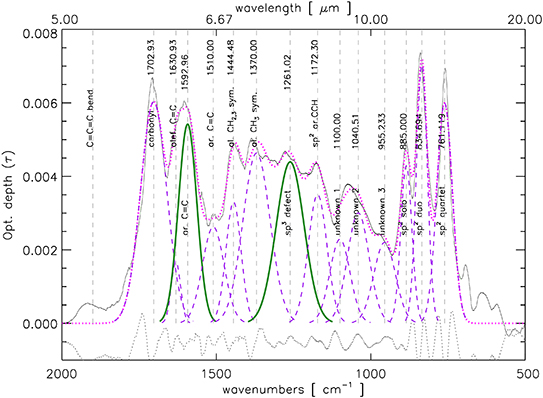}}\\ \vspace{-1em}
\subfloat[ 30 mm ]{\includegraphics[width = 54mm]{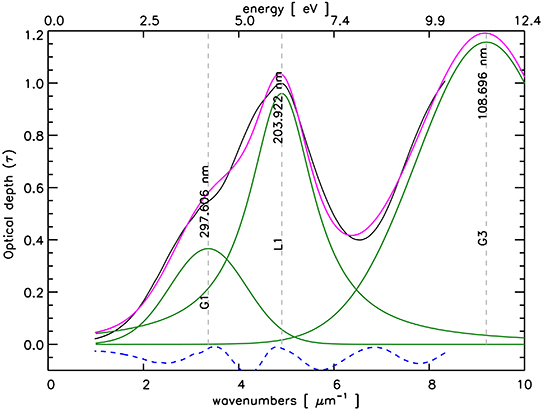}} \hspace{0.5cm}
\subfloat[ 30 mm ]{\includegraphics[width = 54mm]{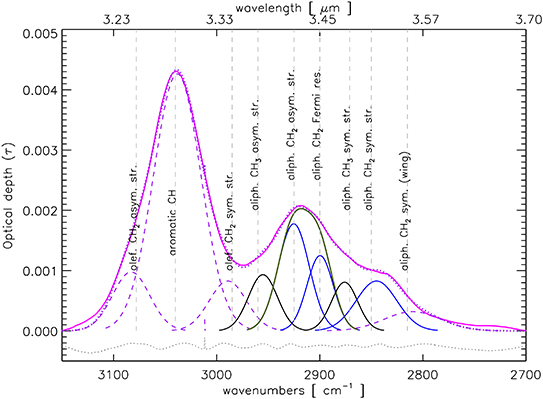}} \hspace{0.5cm}
\subfloat[ 30 mm ]{\includegraphics[width = 54mm]{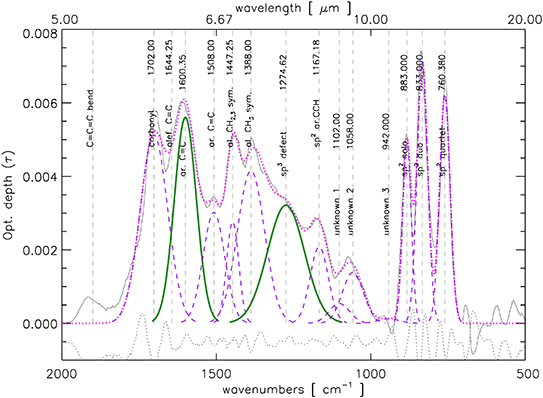}}\\ \vspace{-1em}
\subfloat[ 39 mm ]{\includegraphics[width = 54mm]{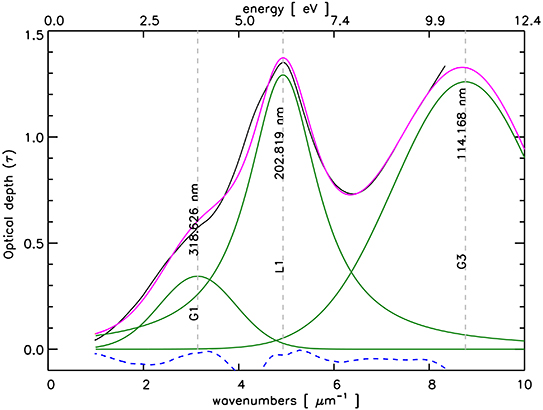}} \hspace{0.5cm}
\subfloat[ 39 mm ]{\includegraphics[width = 54mm]{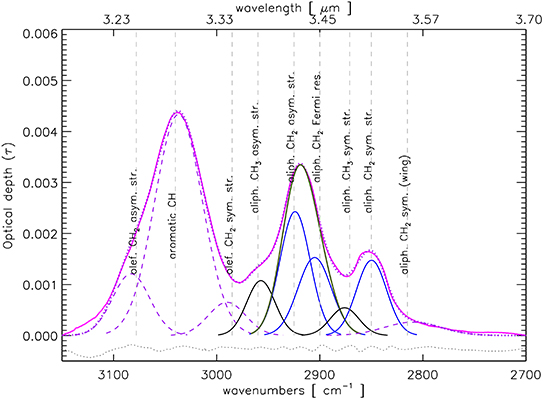}} \hspace{0.5cm}
\subfloat[ 39 mm ]{\includegraphics[width = 54mm]{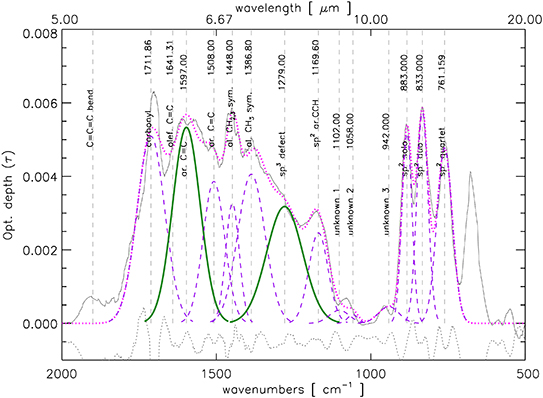}}
\end{figure*}

\newpage

\begin{figure*}
\centering
\label{fig:a7}
\caption{UV-VUV and mid-infrared spectral deconvolution for  an a-C:H produced with C$_7$H$_8$. }
\vspace{1em}
\subfloat[ VUV ]{\includegraphics[width = 54mm]{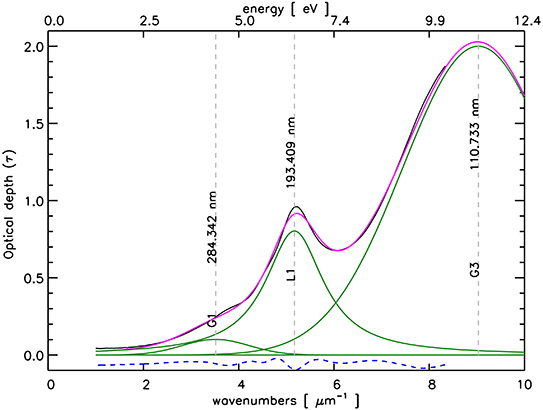}} \hspace{0.5cm}
\subfloat[ IR ]{\includegraphics[width = 54mm]{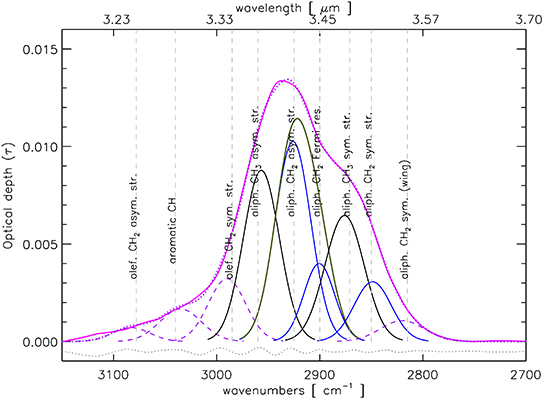}} \hspace{0.5cm}
\subfloat[ IR ]{\includegraphics[width = 54mm]{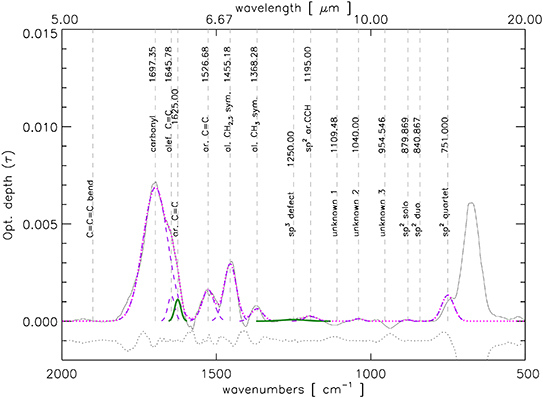}}
\end{figure*}

\newpage

\begin{landscape}

\begin{table}[]
\centering
\caption{UV and VUV fitting parameters}
\label{Table1}
\begin{tabular}{cccccccccccccccccc}
\toprule
flame & d & E$_{g}$ & $\omega_c$(L1) & $\omega_c$(G1) & $\omega_c$(G2) & $\omega_c$(G3)  & $\gamma$(L1) & $\gamma$(G1) & $\gamma$(G2) & $\gamma$(G3) & A(L1) & A(G1) & A(G2) & A(G3) \\
C/O & {[} mm {]} & {[}  eV  {]} & {[} cm$^{-1}$ {]} & {[} cm$^{-1}$ {]} & {[} cm$^{-1}$ {]} & {[} cm$^{-1}$ {]} & {[} cm$^{-1}$ {]} & {[} cm$^{-1}$ {]} & {[} cm$^{-1}$ {]} & {[} cm$^{-1}$ {]} & {[} cm$^{-1}$ {]} & {[} cm$^{-1}$ {]} & {[} cm$^{-1}$ {]} & {[} cm$^{-1}$ {]} \\
\midrule
 & 20 & 0.82 $\pm$ 0.20 & 40123 & NA & 56733 & 91010 & 25118 & NA & 9000 & 43128 & 24725 & NA & 1720 & 50729  \\
0.82 & 21 & 0.95 $\pm$ 0.21  & 39800 & NA & 55122 & 91437 & 21094 & NA & 8000 & 33800 & 2997 & NA & 300 & 7500 \\
 & 22 & 0.79 $\pm$ 0.25  & 40041 & NA & 52988 & 90470 & 22136 & NA & 12277 & 32212 & 6064 & NA & 638 & 8196  \\
 & 30 & 0.67 $\pm$ 0.30   & 39248 & NA & 55319 & 94136 & 21711 & NA & 10000 & 40317 & 14365 & NA & 1110 & 18455  \\
 &  &  &  &  &  &  &  &  &  &  &  &  &  &  \\
 & 18 & 1.44 $\pm$ 0.05 & 48393 & 34498 & NA & 95960 & 17438 & 16385 & NA & 36810 & 12782 & 2228 & NA & 50100  \\
 & 21 & 1.11 $\pm$ 0.08 & 45862 & 30939 & NA & 97122 & 18124 & 13301 & NA & 47060  & 16240 & 1502 & NA & 39132 \\
1 & 25 & 0.84  $\pm$ 0.15 & 42653 & NA & 53663 & 91242 & 25958 & NA & 15091 & 39000 & 11118 & NA & 964 & 10260   \\
 & 30 & 0.72  $\pm$ 0.21  & 39971 & NA & 51638 & 95063 & 24267 & NA & 12162 & 35765 & 12721 & NA & 993 & 17451 \\
 & 39 & 0.61 $\pm$ 0.25  & 40180 & NA & 48915 & 90115 & 26774 & NA & 9000 & 34401 & 6580 & NA & 200 & 8000  \\
 &  &  &  &  &  &  &  &  &  &  &  &  &  &  \\
 & 14 & 1.46 $\pm$ 0.02  & 48509 & 34165 & NA & 91815  & 15879 & 11705 & NA & 38870 & 17806 & 2000 & NA & 35240  \\
1.05 & 18 & 1.14 $\pm$ 0.05  & 46389 & 30894 & NA & 91843 & 18823 & 14482 & NA & 36160 & 19742 & 2187 & NA & 34210  \\
 & 22 & 0.67  $\pm$ 0.16  & 41467 & NA & 52544 & 92021  & 25916 & NA & 10000 & 45067 & 18428 & NA & 799 & 17124  \\
 &  &  &  &  &  &  &  &  &  &  &  &  &  &  \\
 & 22 & 0.74 $\pm$ 0.09  & 42273 & NA & 53000 & 91050 & 26827 & 19165 & 19000 & 42000 & NA & 2999 & 48470  \\
 & 24 & 0.65 $\pm$ 0.16  & 40800 & NA & 53000 & 91763 & 26000 & NA & 13202 & 19000 & 31000 & NA & 2012 & 37420 \\
 & 26 & 0.26 $\pm$ 0.10  & 39900 & NA & 53000 & 91179 & 26500 & NA & 8000 & 19000 & 50000 & NA & 1610 & 45893  \\
 & 28 & 0.39 $\pm$ 0.08  & 39583 & NA & 53000 & 91050 & 27000 & NA & 7254 & 17436 & 46147 & NA & 759 & 38380  \\
1.05 & 30 & 0.20 $\pm$ 0.05  & 39426 & NA & 53000 & 91111 & 24000 & NA & 8000 & 19000 & 50000 & NA & 1544 & 52000 \\
 & 32 & 0.30 $\pm$ 0.11  & 38886 & NA & 51692 & 91050 & 26000 & NA & 10000 & 19000 & 39000 & NA & 1367 & 35000 \\
 & 34 & 0.01 $\pm$ 0.01  & 39348 & NA & 50000 & 91050 & 23000 & NA & 8000 & 19000 & 52000 & NA & 2000 & 45000 \\
 & 36 & 0.18 $\pm$ 0.05 & 38141 & NA & 52000 & 91050 & 23000 & NA & 7000 & 22000 & 18000 & NA & 500 & 51700 \\
 & 42 & 0.19 $\pm$ 0.05  & 37995 & NA & 52000 & 91050 & 22000 & NA & 14011 & 22000 & 26000 & NA & 1749 & 51700 \\
 & 50 & 0.05 $\pm$ 0.02  & 37995 & NA & 50000 & 91050 & 21000 & NA & 7999 & 21000 & 29000 & NA & 600 & 49350 \\
 &  &  &  &  &  &  &  &  &  &  &  &  &  &  \\
 & 25 & 1.60 $\pm$ 0.01 & 48873 & 34283 & NA & 96059  & 16626 & 11415 & NA & 44000  & 18452 & 2000 & NA & 42270  \\
 & 27 & 1.51 $\pm$ 0.01 & 49628 & 33000 & NA & 91206  & 20000 & 9900 & NA & 35360 & 14997 & 1200 & NA & 36150  \\
1.3 & 30 & 1.41 $\pm$ 0.01  & 48700 & 32769 & NA & 93890 & 17578 & 12228 & NA & 44000 & 26098 & 3000 & NA & 44990  \\
 & 39 & 1.00 $\pm$ 0.03 & 49305 & 31021 & NA & 96073 & 19126 & 13066 & NA & 49370 & 37108 & 3000 & NA & 69270 \\
 &  &  &  &  &  &  &  &  &  &  &  &  &  &  \\
a-C:H & NA & 1.95 $\pm$ 0.10 & 52219 & 38583 & NA & 91175  & 11731 & 16464 & NA & 39323 & 12527 & 966 & NA & 26057 \\
\bottomrule
\end{tabular}
\end{table}

\end{landscape}

\newpage

\begin{landscape}

\begin{table}[]
\centering
\caption{Infrared fitting parameters}
\label{Table2}
\begin{tabular}{cccccccccccccc}
\toprule
 &  & \multicolumn{3}{l}{band positions} &  & \multicolumn{8}{c}{integrated vibrational modes} \\
flame & distance & $\omega_c$(C=C) & $\omega_c$sp$^3$ & $\Delta$(CC - sp3) &  & CH$_{arom.}$ & CH$_{3,aliph.}$ & CH$_{2,aliph.}$ & CH$_{3,Fermi.}$ & C=O & C=C & sp$^3$ & sp$^2$ solo \\
C/O & {[} mm {]} & {[} cm$^{-1}$ {]} & {[} cm$^{-1}$ {]} & {[} cm$^{-1}$ {]} &  & {[} cm$^{-1}$ {]} & {[} cm$^{-1}$ {]} & {[} cm$^{-1}$ {]} & {[} cm$^{-1}$ {]} & {[} cm$^{-1}$ {]} & {[} cm$^{-1}$ {]} & {[} cm$^{-1}$ {]} & {[} cm$^{-1}$ {]} \\
\midrule
 & 20 & 1595 & 1247 & 348 &  & 0.05 & 0.10 & 0.10 & 0.10 & 0.63 & 0.41 & 1.97 & 0.40 \\
0.82 & 21 & 1593 & 1243 & 350 &  & 0.03 & 0.06 & 0.07 & 0.08 & 0.48 & 0.33 & 0.98 & 0.13 \\
 & 22 & 1593 & 1242 & 351 &  & 0.04 & 0.09 & 0.07 & 0.06 & 0.53 & 0.39 & 1.86 & 0.34 \\
 & 30 & 1590 & 1232 & 358 &  & 0.02 & 0.04 & 0.06 & 0.08 & 0.41 & 0.38 & 1.67 & 0.26 \\
 &  &  &  &  &  &  &  &  &  &  &  &  &  \\
 & 18 & 1599 & 1280 & 319 &  & 0.03 & 0.06 & 0.04 & 0.02 & 0.18 & 0.08 & 0.15 & NA \\
 & 21 & 1601 & 1280 & 321 &  & 0.03 & 0.06 & 0.05 & 0.04 & 0.18 & 0.14 & 0.18 & NA \\
1 & 25 & 1592 & 1262 & 330 &  & 0.37 & 0.28 & 0.35 & 0.24 & 2.48 & 2.86 & 4.58 & 0.56 \\
 & 30 & 1587 & 1242 & 346 &  & 0.25 & 0.15 & 0.16 & 0.19 & 1.44 & 2.33 & 4.70 & 0.23 \\
 & 39 & 1582 & 1219 & 363 &  & 0.32 & 0.01 & 0.06 & 0.06 & 0.72 & 2.49 & 6.79 & 0.50 \\
 &  &  &  &  &  &  &  &  &  &  &  &  &  \\
 & 14 & 1595 & 1274 & 321 &  & 0.07 & 0.04 & 0.04 & 0.02 & 0.33 & 0.20 & 0.27 & NA \\
1.05 & 18 & 1602 & 1275 & 327 &  & 0.03 & 0.03 & 0.05 & 0.02 & 0.14 & 0.16 & 0.45 & NA \\
 & 22 & 1593 & 1269 & 324 &  & 0.01 & 0.02 & 0.03 & 0.02 & 0.07 & 0.14 & 0.25 & NA \\
 &  &  &  &  &  &  &  &  &  &  &  &  &  \\
 & 22 & 1596 & 1262 & 334 &  & 0.23 & 0.19 & 0.25 & 0.16 & 2.03 & 2.02 & 4.27 & 0.09 \\
 & 24 & 1592 & 1251 & 341 &  & 0.23 & 0.20 & 0.21 & 0.26 & 2.70 & 3.20 & 6.80 & 0.11 \\
 & 26 & 1588 & 1240 & 349 &  & 0.15 & 0.08 & 0.13 & 0.19 & 2.60 & 3.53 & 8.86 & 0.23 \\
 & 28 & 1587 & 1230 & 357 &  & 0.11 & 0.04 & 0.07 & 0.10 & 1.42 & 2.69 & 6.72 & 0.16 \\
1.05 & 30 & 1586 & 1225 & 360 &  & 0.11 & 0.03 & 0.06 & 0.08 & 1.03 & 2.47 & 6.55 & 0.17 \\
 & 32 & 1582 & 1217 & 365 &  & 0.11 & 0.03 & 0.08 & 0.06 & 0.73 & 2.30 & 7.06 & 0.32 \\
 & 34 & 1579 & 1213 & 366 &  & 0.12 & 0.03 & 0.12 & 0.12 & 1.07 & 2.19 & 9.51 & 1.39 \\
 & 36 & 1579 & 1195 & 384 &  & 0.22 & 0.31 & 0.67 & 0.30 & 1.71 & 3.42 & 16.14 & 1.80 \\
 & 42 & 1579 & 1191 & 389 &  & 0.12 & 0.04 & 0.10 & 0.11 & 1.72 & 2.58 & 9.08 & 0.42 \\
 & 50 & 1575 & 1179 & 396 &  & 0.15 & 0.04 & 0.12 & 0.16 & 2.25 & 2.47 & 9.55 & 0.47 \\
 &  &  &  &  &  &  &  &  &  &  &  &  &  \\
 & 25 & 1590 & 1267 & 323 &  & 0.11 & 0.01 & 0.01 & 0.00 & 0.73 & 0.45 & 0.49 & 0.20 \\
 & 27 & 1593 & 1261 & 332 &  & 0.22 & 0.03 & 0.07 & 0.05 & 0.64 & 0.40 & 0.50 & 0.19 \\
1.3 & 30 & 1600 & 1275 & 326 &  & 0.25 & 0.03 & 0.07 & 0.04 & 0.57 & 0.50 & 0.48 & 0.17 \\
 & 33 & 1598 & 1273 & 325 &  & 0.57 & 0.06 & 0.12 & 0.10 & 0.69 & 0.83 & 0.61 & 0.48 \\
 & 36 & 1601 & 1276 & 325 &  & 0.59 & 0.05 & 0.21 & 0.08 & 0.83 & 1.22 & 0.90 & 0.61 \\
 & 39 & 1597 & 1279 & 318 &  & 0.25 & 0.04 & 0.09 & 0.06 & 0.57 & 0.60 & 0.46 & 0.21 \\
 &  &  &  &  &  &  &  &  &  &  &  &  &  \\
a-C:H & NA & 1625 & NA & NA &  & 0.08 & 0.38 & 0.44 & 0.15 & 0.69 & 0.03 & NA & NA \\
\bottomrule

\end{tabular}
\end{table}

\end{landscape}

\end{appendix}

\end{document}